\documentclass[aps,amsmath,amssymb,reprint, double column,prl, showkeys]{revtex4-2}

\usepackage{graphicx,epsfig}
\usepackage{amssymb}
\usepackage{amsmath}
\usepackage{bm}
\usepackage{textcomp}
\usepackage{color}
\usepackage{mathtools}
\usepackage[dvipsnames]{xcolor}

\usepackage{soul}

\begin{document}
\setcounter{page}{1}

\title[]{Fractional quantum Hall effect energy gaps: role of electron layer thickness}
\author{K. A. \surname{Villegas Rosales}}
\author{P. T. \surname{Madathil}}
\author{Y. J. \surname{Chung}}
\author{L. N. \surname{Pfeiffer}}
\author{K. W. \surname{West}}
\author{K. W. \surname{Baldwin}}
\author{M. \surname{Shayegan}}
\affiliation{Department of Electrical Engineering, Princeton University, Princeton, New Jersey 08544, USA}

\date{\today}

\begin{abstract}

The fractional quantum Hall effect (FQHE) stands as a quintessential manifestation of an interacting two-dimensional electron system. One of FQHE's most fundamental characteristics is the energy gap separating the incompressible ground state from its excitations. Yet, despite nearly four decades of investigations, a quantitative agreement between the theoretically calculated and experimentally measured energy gaps is lacking. Here we report a quantitative comparison between the measured energy gaps and the available theoretical calculations that take into account the role of finite layer thickness and Landau level mixing. Our systematic experimental study of the FQHE energy gaps uses very high-quality two-dimensional electron systems confined to GaAs quantum wells with varying well widths. All the measured energy gaps fall bellow the calculations, but as the electron layer thickness increases, the results of experiments and calculations come closer. Accounting for the role of disorder in a phenomenological manner, we find the measured energy gaps to be in reasonable quantitative agreement with calculations, although some discrepancies remain.

\end{abstract}

\maketitle  

The fractional quantum Hall effect (FQHE) \cite{Tsui.PRL.1982,Jain.CFbook.2007,Perspectives.Pinczuk.Das.Sarma,Halperin.Fractionalbook.2020}, is one of the most celebrated many-body phenomena in condensed matter physics. It has been under study for nearly 40 years and has influenced other fields of physics beyond condensed matter \cite{Klitzing.40years.2020}.  It is observed in clean (low-disorder) two-dimensional electron systems (2DESs) when cooled to low temperatures and exposed to a large perpendicular magnetic field so that the thermal and kinetic (Fermi) energies are quenched, and electron-electron interaction dominates. The FQHE signals the formation of an incompressible liquid that has a quantized Hall resistance and flows without dissipation in the limit of zero temperature. It is also the first discovered topological state whose emergence requires interaction \cite{Haldane.NobelLecture.2017}, and some FQHE states are indeed considered prime platforms for topological quantum computing \cite{Nayak.RMP.2008,Gul.inducedSC.2020,Mong.TopologicalComputing.2014}. 

The strongest and most basic FQHE is observed at Landau level (LL) filling factor $\nu=1/3$ \cite{Tsui.PRL.1982}. The physical properties of this state have been of continued fascination and research since its discovery \cite{Jain.CFbook.2007,Perspectives.Pinczuk.Das.Sarma,Halperin.Fractionalbook.2020,Nakamura.NatPhys.2019,Nakamura.NatPhys.2020,Bartolomei.Science.2020,dePiccioto.Nature.1997,Saminadayar.PRL.1997,Morf.PRB.1986,Haldane.PRL.1985,Girvin.PRL.1985,Boebinger.PRL.1985,Willett.PRB.1988,Du.PRL.1993,Pan.PRL.2020,Zhang.PRB.1986,Morf.PRB.2002,Melik-Alaverdian.PRB.1995,Park.Activation.1999,Yoshioka.JPSJ.1984,Yoshioka.SurfScience.1986,Gold.PRB.1987,Platzman.PRB.1989,Shayegan.PRL.1990,He.PRB.1990,Halonen.PRB.1993}. Theoretical calculations predict an energy gap $^{1/3}\Delta=0.10E_{C}$ for an ``ideal" 2DES with zero layer thickness, no disorder, and an infinite separation between the LLs so that the ground state is formed entirely within the lowest LL \cite{Morf.PRB.1986,Haldane.PRL.1985,Girvin.PRL.1985,Jain.CFbook.2007}. The parameter  $E_{C}=\textit{e}^{2}$/$4\pi\epsilon_{0}\epsilon$\textit{l}$_{B}$ is the Coulomb energy, where $l_{B}=\sqrt{\hbar/\textit{e}B}$ is the magnetic length. The energy gaps measured in realistic, experimental samples, however, are smaller than the ideal value because the finite (non-zero) thickness of the electron layer, the proximity of the higher LLs, i.e., the ensuing LL mixing (LLM), and the ubiquitous disorder all tend to lower the energy gaps. Assessing the role of these factors, and a quantitative comparison of the measured and calculated $^{1/3}\Delta$ has been of interest for a long time. Experimentally, with improvements in sample quality, the measured energy gaps have generally increased \cite{Boebinger.PRL.1985,Willett.PRB.1988,Du.PRL.1993,Pan.PRL.2020}. On the theoretical side, realistic factors such as electron layer thickness \cite{Zhang.PRB.1986,Morf.PRB.2002,Melik-Alaverdian.PRB.1995,Park.Activation.1999}, LLM \cite{Yoshioka.JPSJ.1984,Yoshioka.SurfScience.1986,Melik-Alaverdian.PRB.1995}, and disorder \cite{Gold.PRB.1987,Platzman.PRB.1989} have been considered to explain the experimental data. Nevertheless, the experimentally deduced gaps still fall bellow what theory predicts.

An important, experimentally controllable parameter that influences $^{1/3}\Delta$ is the thickness ($\tilde w$) of the electron layer. Despite its fundamental importance, systematic experimental measurements of $^{1/3}\Delta$ as a function of $\tilde w$ have been very scarce. Shayegan \textit{et al.} \cite{Shayegan.PRL.1990} performed a study of $^{1/3}\Delta$ as a function of $\tilde w$ for an electron system confined to an AlGaAs quantum well (QW) with a parabolic potential profile. As the electron density ($n$) is varied in this system using gate bias, $\tilde w$ varies also. The measurements provided clear evidence that $^{1/3}\Delta$ decreases as $\tilde w$ increases, and eventually vanishes for sufficiently large values of $\tilde w$. Subsequent calculations \cite{He.PRB.1990,Halonen.PRB.1993} corroborated the experimental results qualitatively. However, there are two notable shortcomings in the measurements of Ref. \cite{Shayegan.PRL.1990}. First, since the 2DES resides in an AlGaAs QW, it suffers significantly from alloy disorder and therefore exhibits reduced gaps. Moreover, as the charge distribution is made wider by increasing $n$, the 2DES experiences more severe alloy disorder because the electron wavefunction spreads more into a region with larger Al alloy fraction. Second, since the experiments are done by tuning $n$, the influence of LLM on $^{1/3}\Delta$ also changes as $\tilde w$ is increased. (The LLM parameter $\kappa$ is defined as the ratio of the Coulomb energy and the LL separation: $\kappa=(e^{2}/4\pi\epsilon_{0}\epsilon l_{B})/(\hbar eB/m^{*})$, where $m^{*}$ is the effective band mass. For a fixed $\nu$, $\kappa \propto 1/\sqrt{n}$.)

\begin{figure*}[t!]
  \centering
    \psfig{file=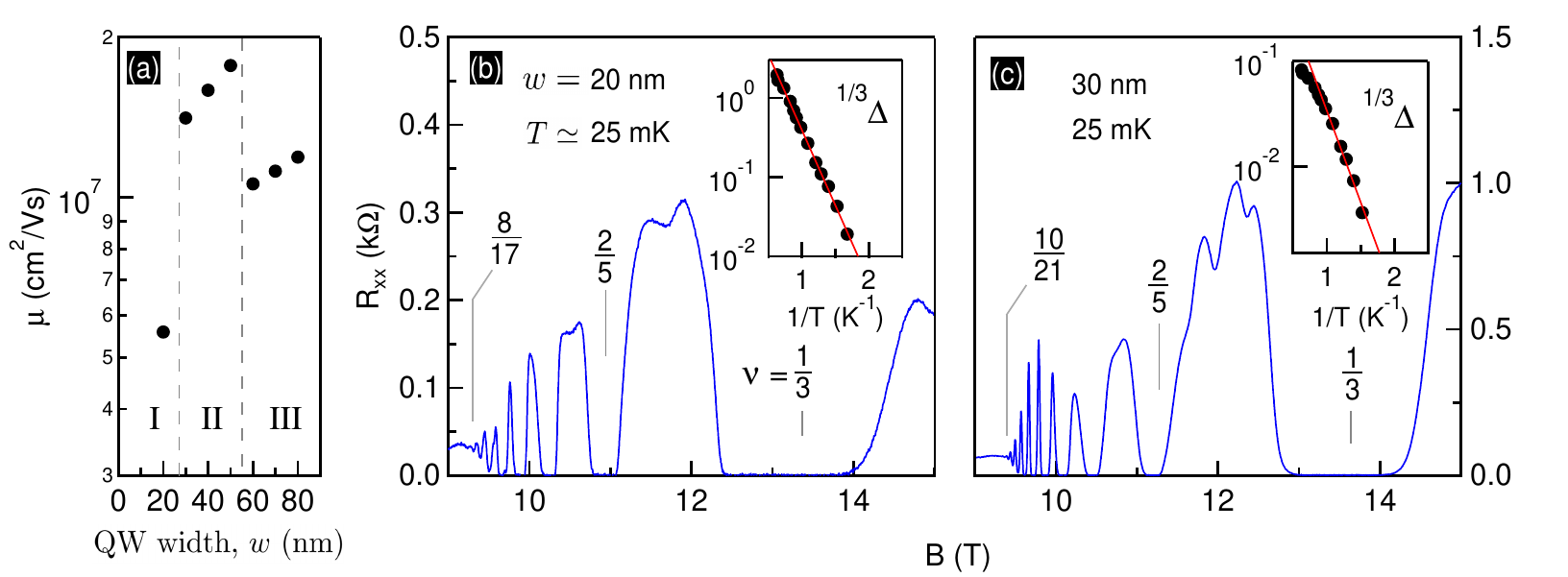, width=1\textwidth}
  \centering
  \caption{\label{transport} 
   (a) Transport mobility ($\mu$) vs. quantum well width ($w$). (b-c) Longitudinal resistance ($R_{xx}$) vs. perpendicular magnetic field (\textit{B}) for GaAs 2DESs with density $\simeq1.1\times10^{11}$ cm$^{-2}$, and $w=$ 20 and 30 nm. The insets show the Arrhenius plots of $R_{xx}$ minimum at $\nu=1/3$ from which we deduce $^{1/3}\Delta$. %C:\Users\kvill\Mirror\5.Maglab\2019_September\analysis_Sat_2nd\analysis_home
  }
  \label{fig:transport}
\end{figure*}

Here we present a systematic experimental study of $^{1/3}\Delta$ vs. $\tilde w$ in extremely high quality 2DESs confined to modulation-doped GaAs QWs grown on GaAs (001) substrates. The QWs are flanked by  150-nm-thick Al$_{0.24}$Ga$_{0.76}$As barriers, and the dopants are placed in doping wells \cite{Chung.PRM.2020}. The 2DESs all have the same density $n\simeq1.1\times10^{11}$ cm$^{-2}$, in order to keep the LLM parameter fixed, and the QW widths (\textit{w}) are varied from $20$ to $80$ nm to change $\tilde w$. We designate each sample by $S_{w}$, e.g., $S_{20}$ refers to the sample with a QW of width $w=20$ nm.  The samples have a $4$ mm $\times$ $4$ mm van der Pauw geometry, with alloyed InSn electrical contacts at the corners and edge midpoints. We used $^{3}$He and  dilution refrigerator systems, and conventional lock-in techniques to obtain magnetoresistance data, and determined $^{1/3}\Delta$ from the activated temperature dependence of the resistance minimum at $\nu=1/3$. In addition, we measured the energy gaps for numerous higher-order FQHE states, and used the gaps to estimate values for disorder in each sample. All our measurements were done without illumination.

The extremely high quality of the samples is illustrated in Fig. 1(a) which shows the low-temperature ($T=0.3$ K) mobility ($\mu$) vs. $w$. Except for $S_{20}$, for all samples $\mu$ exceeds $1\times 10^{7}$ cm$^{2}$/Vs despite the relatively small density. The enhancement in mobility with increasing $w$ seen up to $w=50$ (regions I and II) is related to the reduced role of interface roughness scattering \cite{Li.SST.2005,Kamburov.APL.2016}. When the 2DES is confined to a wider QW, the charge distribution penetrates the flanking Al$_{0.24}$Ga$_{0.76}$As spacers to a smaller extent and the 2DES effectively experiences less alloy disorder. Once $w$ exceeds 50 nm (region III in Fig. 1(a)), the 2DES starts to occupy the second electric sub-band, and the mobility drops precipitously because of the additional inter-sub-band scattering \cite{Stormer.SSC.1982}.

In Figs. 1(b-c) we show the longitudinal resistance ($R_{xx}$) vs. $B$ for $S_{20}$ and $S_{30}$ at \textit{T} $\simeq25$ mK. Numerous minima observed in $R_{xx}$ at fractional fillings $\nu=p/(2p+1)$ attest to a plethora of FQHE states and the high quality of the samples. For $S_{20}$ (Fig. 1(b)), we observe minima up to $\nu=8/17$ ($p=8$), while $S_{30}$ (Fig. 1(c)) exhibits FQHE minima up to $\nu=10/21$ ($p=10$). This is consistent with the higher mobility of $S_{30}$. Data for samples with larger $w$, presented in the Supplementary Material (SM) \cite{SM}, also show FQHE minima up to $p=10$. In Figs. 1(b-c) insets we show Arrhenius plots of $R_{xx}$ at $\nu=1/3$ vs. 1/$T$, from which we determine $^{1/3}\Delta$ using fits of the form: $R_{xx}\propto \exp{(- ^{1/3}\Delta/2T)}$. For $S_{20}$ and $S_{30}$, $^{1/3}\Delta$ are ($8.7\pm0.1$) and ($8.0\pm0.4$) K, respectively. It is noteworthy that, despite its higher mobility and quality, $S_{30}$ has the smaller $^{1/3}\Delta$, already hinting that a larger electron layer thickness leads to a smaller $^{1/3}\Delta$.

\begin{figure}[t!]
  \centering
    \psfig{file=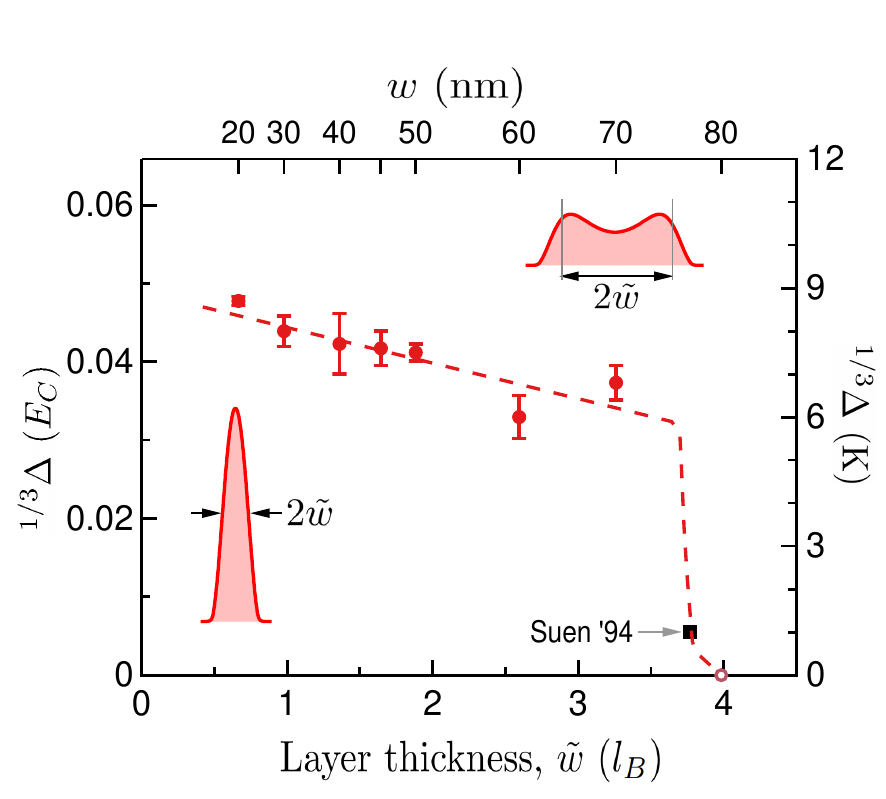, width=0.48\textwidth}
  \centering
  \caption{\label{Ip} 
$^{1/3}\Delta$ (in units of Coulomb energy, $E_{C}$) vs. effective layer thickness $\tilde w$ (in units of magnetic length, $l_{B}$); in our samples $l_{B}\simeq7.1$ nm at $\nu=1/3$,. We show the charge distributions (from self-consistent calculations) for $w=20$ and $70$ nm as insets; $\tilde w$ is defined as the standard deviation of the charge distribution from its center. The black symbol is $^{1/3}\Delta$ by Suen \textit{et al.} \cite{Suen.PRL.b.1994} for a 2DES with similar density to ours and $w=77$ nm. For $w=80$ nm, we find an insulating phase instead of a FQHE and represent it with an open circle.
  }
  \label{fig:Ip}
\end{figure}

Figure 2 presents $^{1/3}\Delta$ (in units $E_{C}$) vs. the effective layer thickness $\tilde w$ (in units of $l_{B}$); each data point is from a different 2DES whose QW width is given in the top axis. We use a Schroedinger-Poisson solver \cite{Schroedinguer.Poisson.solver} to calculate the charge distribution in each QW self-consistently, and define $\tilde w$ as the standard deviation of the charge distribution from its center. Examples are shown in Fig. 2 insets for $w=20$ and $70$ nm. As seen in Fig. 2, $^{1/3}\Delta$ decreases with increasing $\tilde w$, qualitatively expected from the softening of the Coulomb interaction in thicker 2DESs \cite{Zhang.PRB.1986,He.PRB.1990,Shayegan.PRL.1990,Morf.PRB.2002,Halonen.PRB.1993,Melik-Alaverdian.PRB.1995,Park.Activation.1999}. Note that the theoretical value for $^{1/3}\Delta$ in the limit of $\tilde w=0$, no LLM, and no disorder, is $0.10E_{C}$, about a factor of two larger than our largest measured $^{1/3}\Delta$.

When $w>70$ nm, the $\nu=1/3$ FQHE becomes very weak quickly and $^{1/3}\Delta$ eventually vanishes. In Fig. 2 we have included a data point at $w=77$ nm from Suen \textit{et al.} \cite{Suen.PRL.b.1994}. For larger $w$, in sample $S_{80}$, we observe an insulating phase instead of a $\nu=1/3$ FQHE. Such an insulating phase appears when the charge distribution in wide QWs becomes predominantly bilayer-like \cite{Suen.PRL.1992,Suen.PRL.b.1994,Manoharan.PRL.1996}. As detailed elsewhere, the insulating phase signals a correlated, $bilayer$ Wigner crystal state \cite{Manoharan.PRL.1996,Hatke.Nat.Comm.2015,Hatke.PRB.2017,Suen.PRL.1992}. In Fig. 2, we represent this vanishing of the FQHE in sample $S_{80}$ by an open circle for $w=80$ nm ($\tilde w\simeq4.0$). We report the energy gaps as a function of $\tilde w$ for FQHE at $\nu=2/3$, $2/5$, and $3/5$ in the SM.

For a quantitative comparison with the results of calculations, in Fig. 3 we plot the measured $^{1/3}\Delta$ vs. $\tilde w$ and, using open squares, we also present the energy gaps calculated in three different studies all of which assume zero disorder. The green and black symbols represent calculations that include only the role of layer thickness \cite{Park.Activation.1999,Morf.PRB.2002}, while the blue symbols include both layer thickness and LLM \cite{Melik-Alaverdian.PRB.1995}. In Fig. 3 we present Ref. \cite{Melik-Alaverdian.PRB.1995} results for LLM parameter $\kappa\simeq0.68$, which is equal to $\kappa$ in our samples; also note that this value of $\kappa$ reduces the calculated  $^{1/3}\Delta$ from $\simeq0.10 E_{C}$ to $\simeq0.09 E_{C}$ at $\tilde w=0$, as seen in Fig. 3. References \cite{Morf.PRB.2002,Park.Activation.1999} computed the Coulomb potential in a local-density approximation for 2DESs confined to QWs, while in Ref. \cite{Melik-Alaverdian.PRB.1995} the Zhang-Das Sarma potential \cite{Zhang.PRB.1986} for 2DESs confined to heterojunctions was used. 

Although there is an overall, qualitative agreement in Fig. 3 between the experimental and theoretical results in that both exhibit a decrease with increasing  $\tilde w$, the experimental data uniformly fall below calculations, with the difference being largest at the smallest $\tilde w$. Since one possible source for the discrepancy is the presence of disorder, it is useful to extract an experimental estimate for the role of disorder in lowering $^{1/3}\Delta$ in our samples. We do this by considering the energy gaps $^{\nu}\Delta$ for higher-order FQHE states at other $\nu$ near $\nu=1/2$. For an ideal 2DES, these gaps are expected to scale as: $^{\nu}\Delta=(C/|2p+1|)E_{C}$ \cite{HLR.PRB.1993,Jain.CFbook.2007}, where $C\simeq0.3$ and $\nu=p/(2p+1)$. Figure 4 displays $^{\nu}\Delta$ vs. (\textit{e}$^{2}$/$4\pi\epsilon_{0}\epsilon$\textit{l}$_{B}$)/$(2p+1)$ (red symbols) for sample $S_{70}$ for $\nu$ up to $8/17$ and $8/15$. In Fig. 4, we also show red lines representing fits to the measured gaps. The magnitude of the negative intercepts of these lines with the $y$-axis provides an estimate of the disorder energy, $\Gamma$ \cite{Du.PRL.1993,Manoharan.PRL.1994,Pan.PRL.2020}; for the data of Fig. 4, we find $\Gamma=(1.2\pm0.2)$ K. Our $\Gamma$ parameter has approximately half the previous values in 2DESs \cite{Du.PRL.1993,Pan.PRL.2020} and 2D hole systems \cite{Manoharan.PRL.1994}, and is about an order of magnitude smaller than in graphene \cite{Polshyn.PRL.2018}, attesting to the extremely high quality of our 2DESs.

\begin{figure}[t!]
  \centering
    \psfig{file=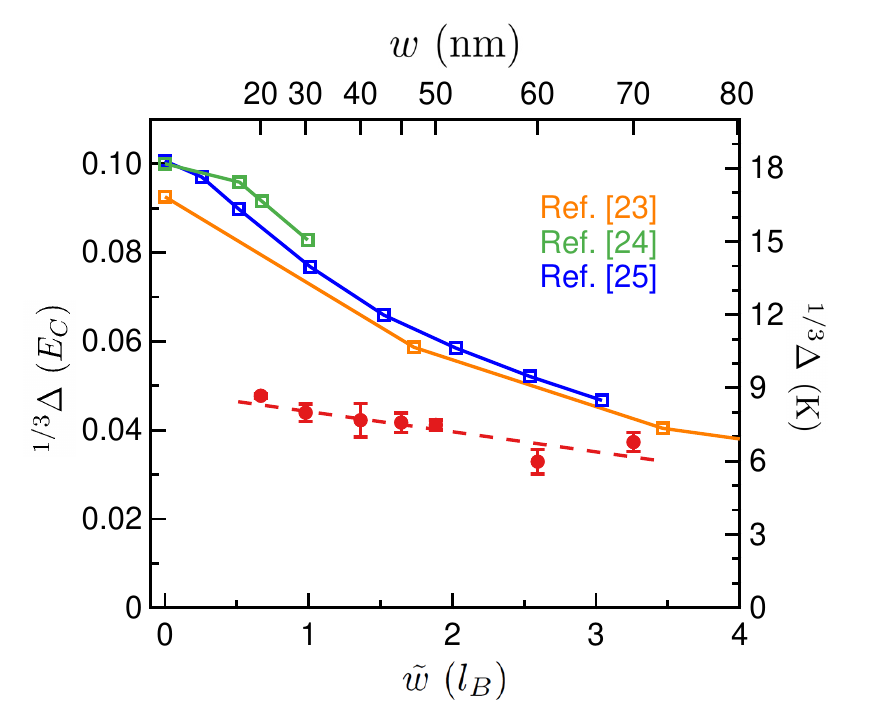, width=0.48\textwidth}
  \centering
  \caption{\label{Ip} 
$^{1/3}\Delta$ vs. $\tilde w$. The open symbols are from theoretical calculations that include the role of finite layer thickness \cite{Park.Activation.1999,Morf.PRB.2002}, and Landau level mixing and finite layer thickness \cite{Melik-Alaverdian.PRB.1995}.
  }
  \label{fig:Ip}
\end{figure}

\begin{figure}[t!]
  \centering
    \psfig{file=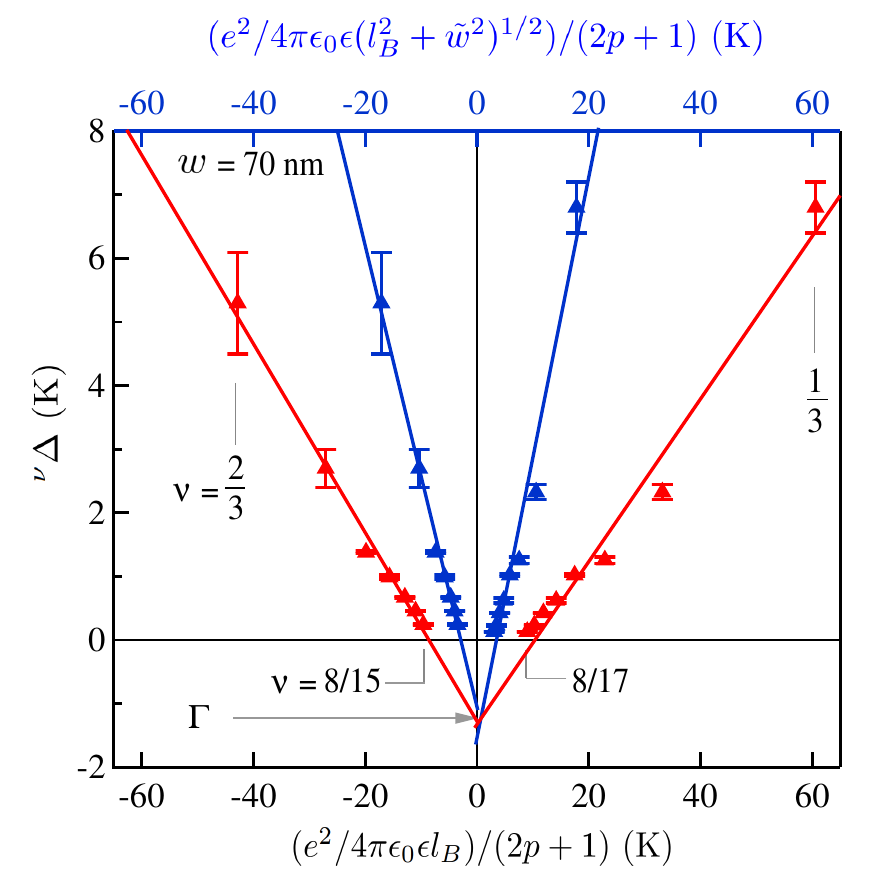, width=0.48\textwidth}
  \centering
  \caption{\label{Ip} 
Red symbols are $^{\nu}\Delta$ vs. $(e^{2}/4\pi\epsilon_{0}\epsilon l_{B})/(2p+1)$, for a 2DES with $w=70$ nm. The $\nu$ range is from $1/3$ to $8/17$, and $2/3$ to $8/15$. The red lines are linear fits to the data. The blue symbols are $^{\nu}\Delta$ vs. the Zhang-Das Sarma energy $(e^{2}/4\pi\epsilon_{0}\epsilon (l_{B}^{2}+\tilde w^{2})^{1/2})/(2p+1)$. The blue lines are linear fits to the blue data points. The red and blue lines have negative intercept values with the $y$-axis that we identify as the phenomenological disorder parameter ($\Gamma$).
  }
  \label{fig:Ip}
\end{figure}

%In Fig. 3, $^{2/3}\Delta\sim$ $^{1/3}\Delta$, which is at odds with the proportionality of $\Delta$ and $e^{2}/\epsilon l_{B}$. In ideal 2DESs, with fixed density, the Coulomb energy is $\sqrt{2}$ times larger at $\nu=1/3$ than at $2/3$, and that leads to $^{1/3}\Delta=\sqrt{2}$ $^{2/3}\Delta$. 
In Fig. 4, the data plotted by red symbols and the fitted lines are not symmetric with respect to  the $y$-axis while theoretically, for an ideal 2DES, they should be symmetric. The asymmetry can be readily attributed to the finite layer thickness of the 2DES: since $l_{B}$ changes with $B$, so does $\tilde w$, implying that the 2DES is effectively wider at higher $B$, or equivalently at smaller $\nu$. To account for the changes in $\tilde w$ as a function $\nu$, we plot our measured gaps vs. the Zhang-Das Sarma energy $(e^{2}/4\pi\epsilon_{0}\epsilon (l_{B}^2+\tilde w^{2})^{1/2})/(2p+1)$ \cite{Zhang.PRB.1986} (top axis and blue symbols in Fig. 4). Because of the reduced Coulomb interaction, the blue symbols cover a smaller range in the $x$-axis compared to the red symbols. The data are now much more symmetric with respect to the $y$-axis, as also clearly seen from the blue lines that are fits through the data points. The average of the intercepts of these lines with the $y$-axis yields $\Gamma=(1.3\pm0.2)$ K, slightly larger than $\Gamma$ deduced from the red lines. Plots similar to Fig. 4 are reported in the SM \cite{SM} for the other samples in our study.

In Fig. 5(b) we summarize the measured $\Gamma$ vs. $\tilde w$, deduced from analyses similar to the one shown in Fig. 4, for all the samples in our study \cite{SM}. For $\Gamma$ we use the average of values deduced from the intercepts of the blue lines with the y-axis in Fig. 4. Note that the measured $\Gamma$ are typically a very small fraction ($\simeq 1\%$) of the Coulomb energy, attesting to the very low disorder in our samples. On the other hand, $\Gamma$ values are not negligible compared to $^{1/3}\Delta$. To account for the sample disorder, we add the measured $\Gamma$ to the measured $^{1/3}\Delta$ for each sample, and show the data in Fig. 5(a). With the added $\Gamma$, the measured energy gaps are overall closer to the theoretical calculations, especially at large layer thicknesses. In the SM \cite{SM}, we display similar data as in Fig. 5(a) for $\nu=2/3$, $2/5$, and $3/5$ FQHE states, and find similar agreements between experiment and theory for 2DESs. 

\begin{figure}[t!]
  \centering
    \psfig{file=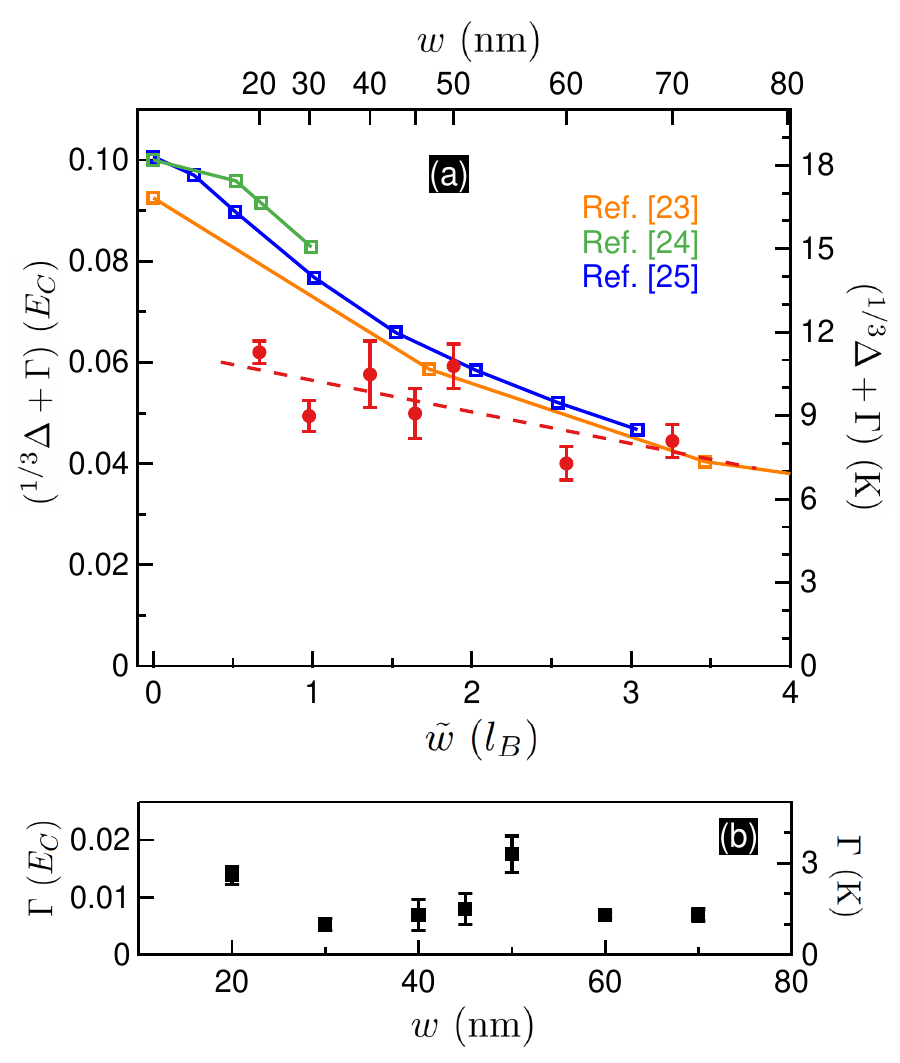, width=0.48\textwidth}
  \centering
  \caption{\label{Ip} 
(a) Closed red symbols are the measured $^{1/3}\Delta$ with the added corresponding $\Gamma$ parameter to account for disorder. (b) The parameter $\Gamma$ vs. $w$.
  }
  \label{fig:Ip}
\end{figure}

To summarize, we present the first systematic experimental study of the dependence of the $\nu=1/3$ FQHE energy gap on the electron layer thicknesses in very high quality 2DESs confined to GaAs QWs. There is a clear decrease of the gap with increasing layer thickness (Fig. 2), qualitatively consistent with the results of available calculations, although some noticeable discrepancies remain, particularly for smaller electron layer thicknesses (Fig. 3). Accounting for the role of disorder, the measured gaps are in better agreement with the calculations (Fig. 5). We close by making two remarks. First, in Fig. 5(a), there is a bit of scatter in the plotted energy gaps. A good portion of this scatter and the error bars in the data points stems from the scatter and error in the deduced disorder parameter $\Gamma$ (Fig. 5(b)). Note also that there appears to be no one-to-one correspondence between the deduced $\Gamma$ and the 2DES mobility at zero magnetic field. For example, data of Fig. 1(a) show the highest mobility for the sample with $w=50$ nm but the same sample exhibits the highest disorder (largest $\Gamma$) in Fig. 5(b). The fact that the experimentally observed strengths of FQHE states and the transport mobilities are not necessarily correlated was first noted in Ref. \cite{Sajoto.PRB.1990}, and remains to be explained rigorously. Second, for a closer, more quantitative comparison of the experimental theoretical results, future calculations based on parameters that better match those of our samples would be welcome. For example, calculations in Ref. \cite{Melik-Alaverdian.PRB.1995}, which do take LLM into account, use a charge distribution which is appropriate for a triangular confinement (GaAs/AlGaAs heterojunction) but not for a QW with a square potential. Also, assessing the role of disorder in modifying the energy gap in realistic samples plays a key role. We hope that our systematic experimental data, taken in extremely high quality 2DESs, would stimulate future theoretical calculations relevant to the parameters of our samples.

\begin{acknowledgments}

We acknowledge support by the NSF (Grant No. DMR 1709076) for measurements, and the NSF (Grant Nos. ECCS 1906253 and No. MRSEC DMR 1420541), and the Gordon and Betty Moore Foundation's EPiQS Initiative (Grant No. GBMF9615 to L.N.P.) for sample synthesis and characterization. This research is funded in part by QuantEmX grants from Institute for Complex Adaptive Matter. A portion of this work was performed at the National High Magnetic Field Laboratory (NHMFL), which is supported by National Science Foundation Cooperative Agreement No. DMR-1644779 and the state of Florida. We thank S. Hannahs, T. Murphy, A. Bangura, G. Jones, and E. Green at NHMFL for technical support. We also thank J. K. Jain for illuminating discussions.

\end{acknowledgments}

%\bibliography{GaAs.13gap}

\end{document}

\setcounter{page}{1}

\title[]{Supplementary Material for ``Fractional quantum Hall effect energy gap: role of electron layer thickness"}
\author{K. A. \surname{Villegas Rosales}}
\author{P. T. \surname{Madathil}}
\author{Y. J. \surname{Chung}}
\author{L. N. \surname{Pfeiffer}}
\author{K. W. \surname{West}}
\author{K. W. \surname{Baldwin}}
\author{M. \surname{Shayegan}}
\affiliation{Department of Electrical Engineering, Princeton University, Princeton, New Jersey 08544, USA}
\date{\today}

\maketitle   

\section{Data at additional quantum well widhts}

In Fig. S1 we show the longitudinal resistance ($R_{xx}$) vs. perpendicular magnetic field ($B$) for a two-dimensional electron system (2DES) confined to a quantum well with width ($w$) of 20 nm, at $25$ mK. The inset displays the Arrhenius plots of $R_{xx}$ at Landau level filling factors $\nu=1/3$ and $\nu=2/3$. We obtain the energy gaps $^{1/3}\Delta$ and $^{2/3}\Delta$ from a fits to the data. Figure S2 presents $^{\nu}\Delta$ vs. (\textit{e}$^{2}$/$4\pi\epsilon_{0}\epsilon$\textit{l}$_{B}$)/$(2p+1)$ (red symbols). To account for the role of electron layer thickness we plot $^{\nu}\Delta$ vs. the Zhang-Das Sarma energy $(e^{2}/4\pi\epsilon_{0}\epsilon (l_{B}^2+\tilde w^{2})^{1/2})/(2p+1)$ (top axis and blue symbols). The red and blue lines are fits to the data. The magnitude of the negative
intercepts of these lines with the $y$-axis provides an estimate of the disorder ($\Gamma$). As seen in Fig. S2, the negative intercepts of the lines for positive and negative $x$-values are different for this sample. To obtain $\Gamma$, we use the average of values deduced from the blue lines in Fig. S2. 

Figures S3 to S14 show magnetoresistance and energy gap data for samples with $w=30$, $40$, $45$, $50$, $60$, and $70$ nm.

\clearpage

\begin{figure*}[t!]
  \centering
    \psfig{file=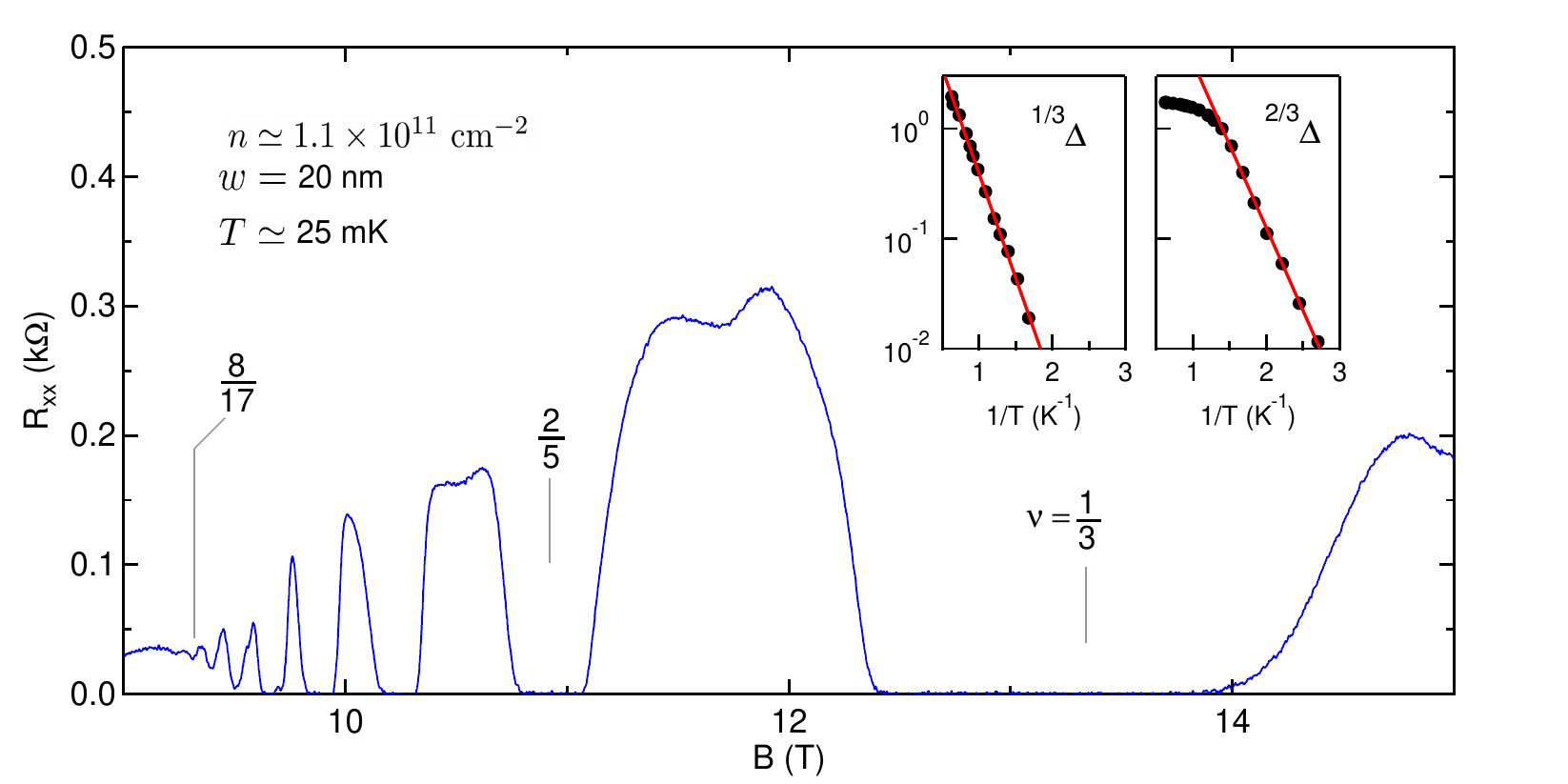, width=1\textwidth}
  \centering
  \caption{\label{transport} 
   $R_{xx}$ vs. \textit{B} for a GaAs 2DES with $w=20$ nm. The insets show the Arrhenius plots of $R_{xx}$ minimum at $\nu=1/3$ and $\nu=2/3$. We fit the experimental data to $R_{xx} \propto \exp{(-^{\nu}\Delta/2T)}$ and deduce energy gaps ($8.7$ $\pm$ $0.1$) K and ($6.9\pm0.1$) K for $^{1/3}\Delta$ and $^{2/3}\Delta$, respectively.
  }
  \label{fig:transport}
\end{figure*}

\begin{figure}[b!]
  \centering
    \psfig{file=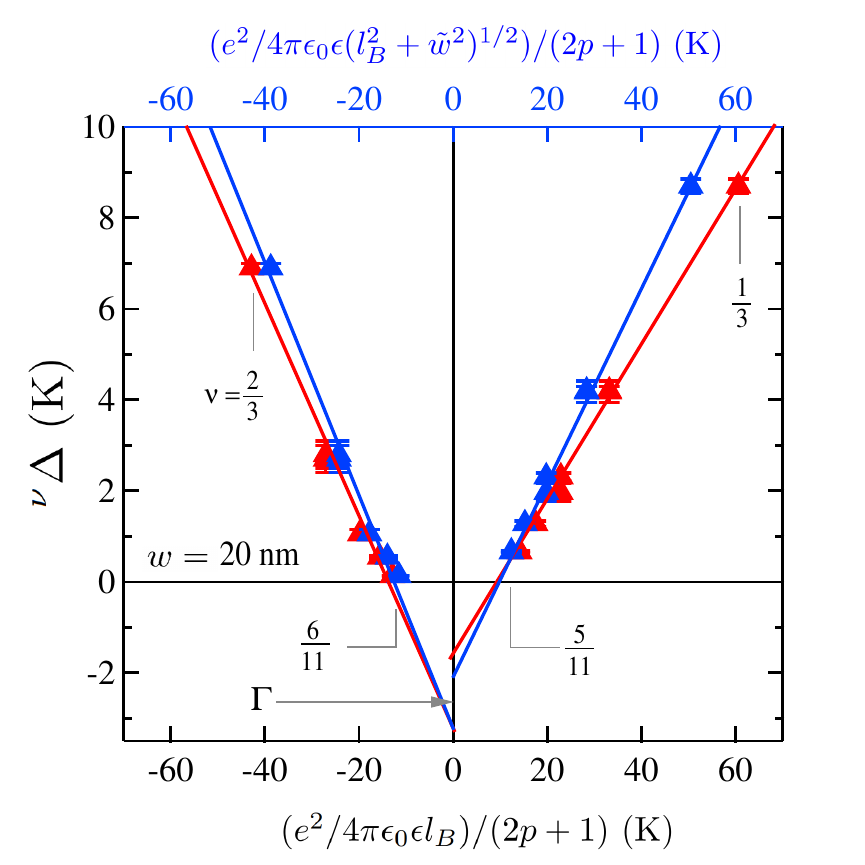, width=0.45\textwidth}
  \centering
  \caption{\label{transport} 
   Red symbols are $^{\nu}\Delta$ vs. $(e^{2}/4\pi\epsilon_{0}\epsilon l_{B})/(2p+1)$, for a 2DES with $w=20$ nm. The blue symbols are $^{\nu}\Delta$ vs. the Zhang-Das Sarma energy $(e^{2}/4\pi\epsilon_{0}\epsilon (l_{B}^{2}+\tilde w^{2})^{1/2})/(2p+1)$. The red and blue lines are linear fits to the data, and their negative intercepts yield an estimate for the disorder paramater $\Gamma=2.6\pm0.3$ K.
  }
  \label{fig:transport}
\end{figure}

\clearpage

\begin{figure*}[h!]
  \centering
    \psfig{file=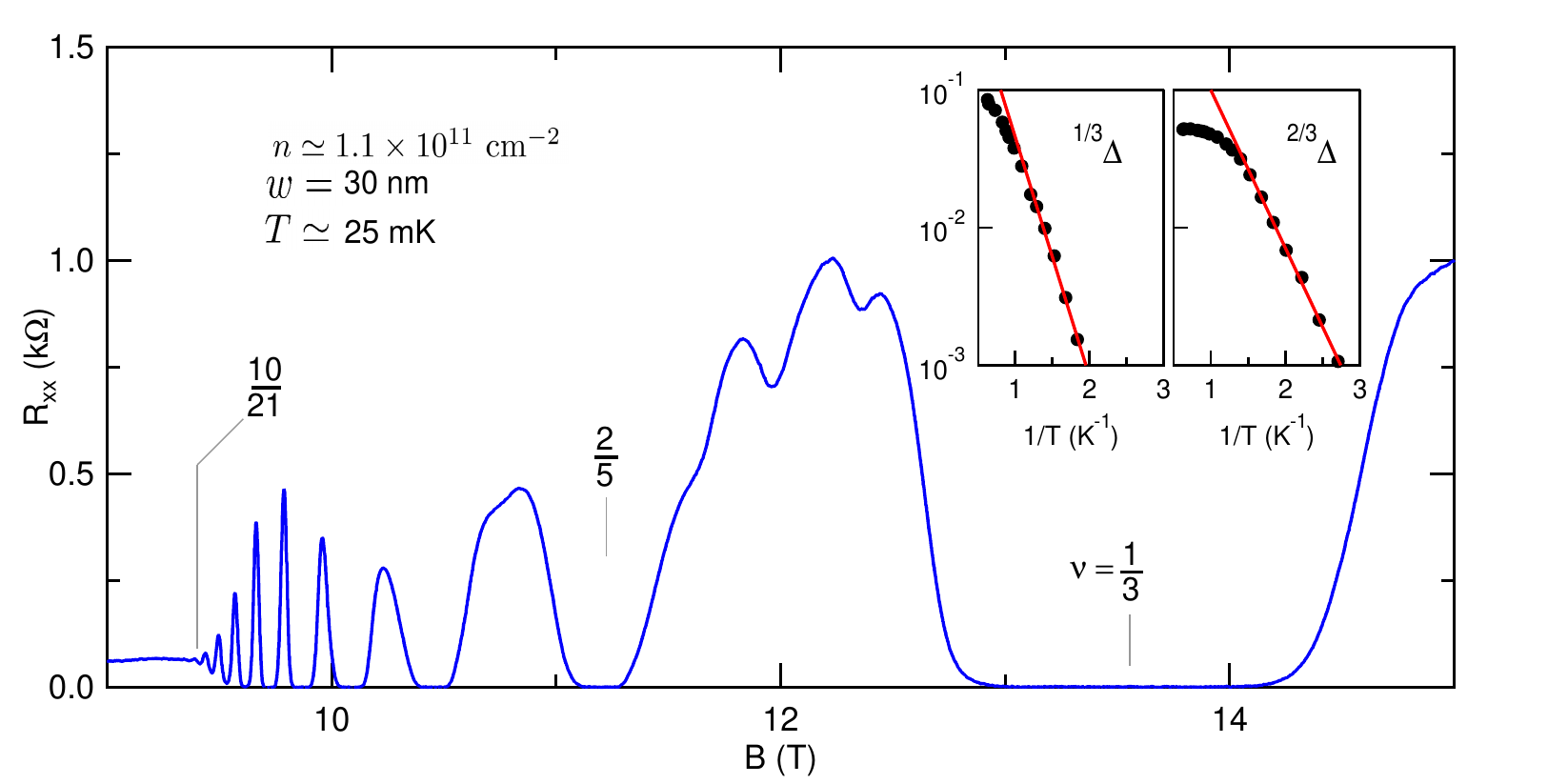, width=1\textwidth}
  \centering
  \caption{\label{transport} 
   $R_{xx}$ vs. \textit{B} for a GaAs 2DES with $w=30$ nm. The deduced $^{1/3}\Delta$ and $^{2/3}\Delta$ are ($8.0$ $\pm$ $0.3$) K and ($5.3\pm0.1$) K, respectively.
  }
  \label{fig:transport}
\end{figure*}

\begin{figure}[h!]
  \centering
    \psfig{file=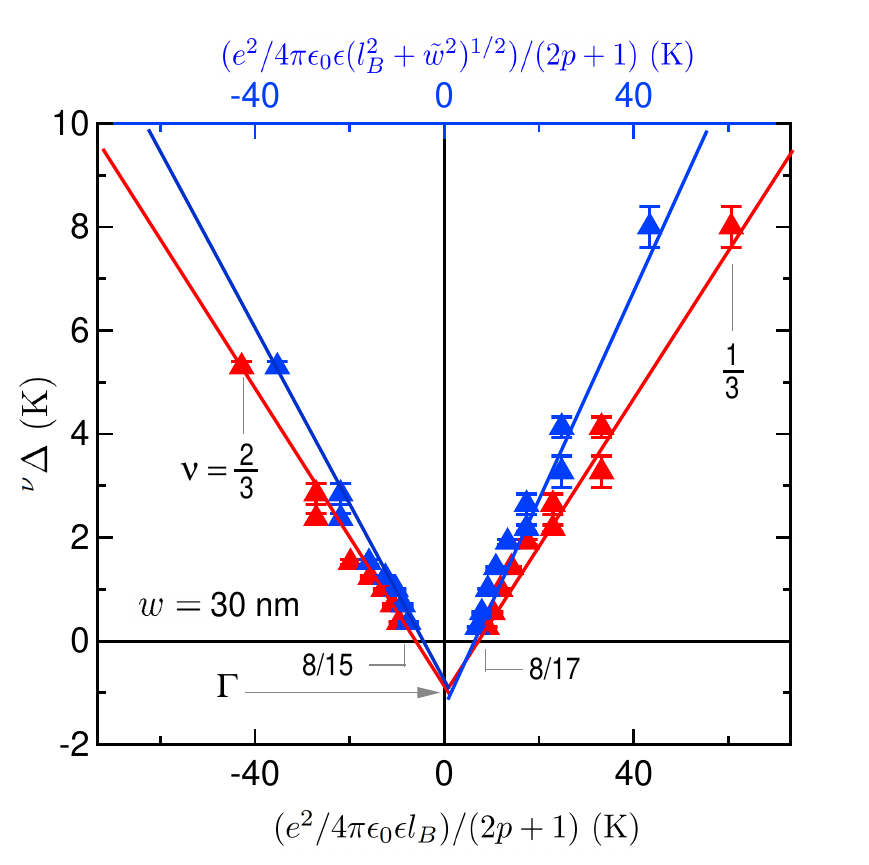, width=0.45\textwidth}
  \centering
  \caption{\label{transport} 
   Red symbols are $^{\nu}\Delta$ vs. $(e^{2}/4\pi\epsilon_{0}\epsilon l_{B})/(2p+1)$, for a 2DES with $w=30$ nm. The blue symbols are $^{\nu}\Delta$ vs. the Zhang-Das Sarma energy $(e^{2}/4\pi\epsilon_{0}\epsilon (l_{B}^{2}+\tilde w^{2})^{1/2})/(2p+1)$. The red and blue lines are linear fits to the data, and $\Gamma=1.0\pm0.2$ K.
  }
  \label{fig:transport}
\end{figure}

\clearpage

\begin{figure*}[h!]
  \centering
    \psfig{file=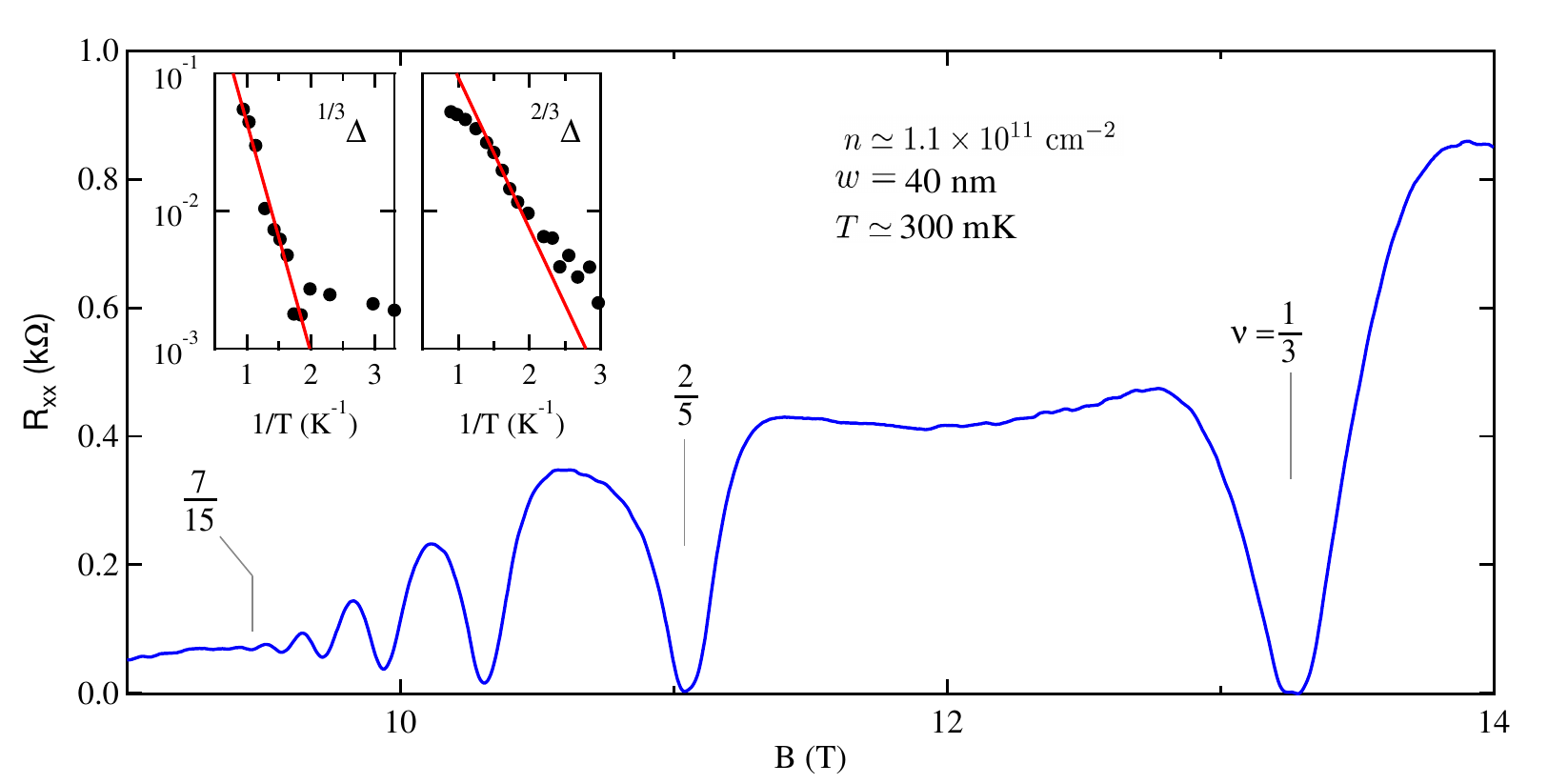, width=1\textwidth}
  \centering
  \caption{\label{transport} 
   $R_{xx}$ vs. \textit{B} for a GaAs 2DES with $w=40$ nm. The deduced $^{1/3}\Delta$ and $^{2/3}\Delta$ are ($7.7$ $\pm$ $0.7$) K and ($5.1\pm0.2$) K.
  }
  \label{fig:transport}
\end{figure*}

\begin{figure}[h!]
  \centering
    \psfig{file=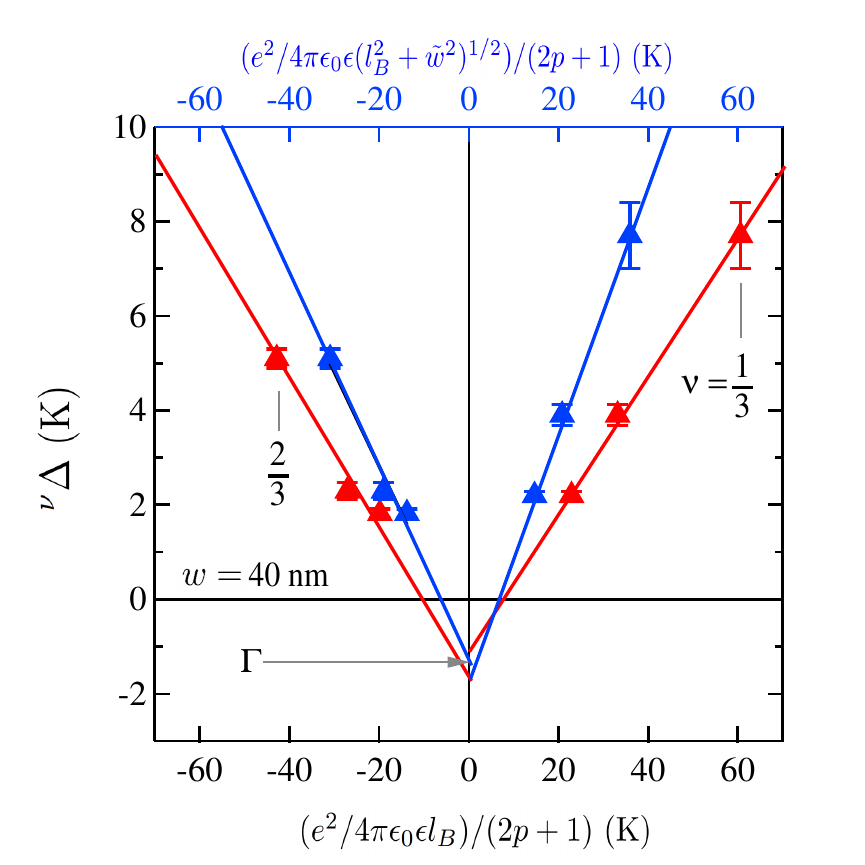, width=0.45\textwidth}
  \centering
  \caption{\label{transport} 
   Red symbols are $^{\nu}\Delta$ vs. $(e^{2}/4\pi\epsilon_{0}\epsilon l_{B})/(2p+1)$, for a 2DES with $w=40$ nm. The blue symbols are $^{\nu}\Delta$ vs. the Zhang-Das Sarma energy $(e^{2}/4\pi\epsilon_{0}\epsilon (l_{B}^{2}+\tilde w^{2})^{1/2})/(2p+1)$. The red and blue lines are linear fits to the data, and $\Gamma=1.3\pm0.5$ K.
  }
  \label{fig:transport}
\end{figure}

\clearpage

\begin{figure*}[h!]
  \centering
    \psfig{file=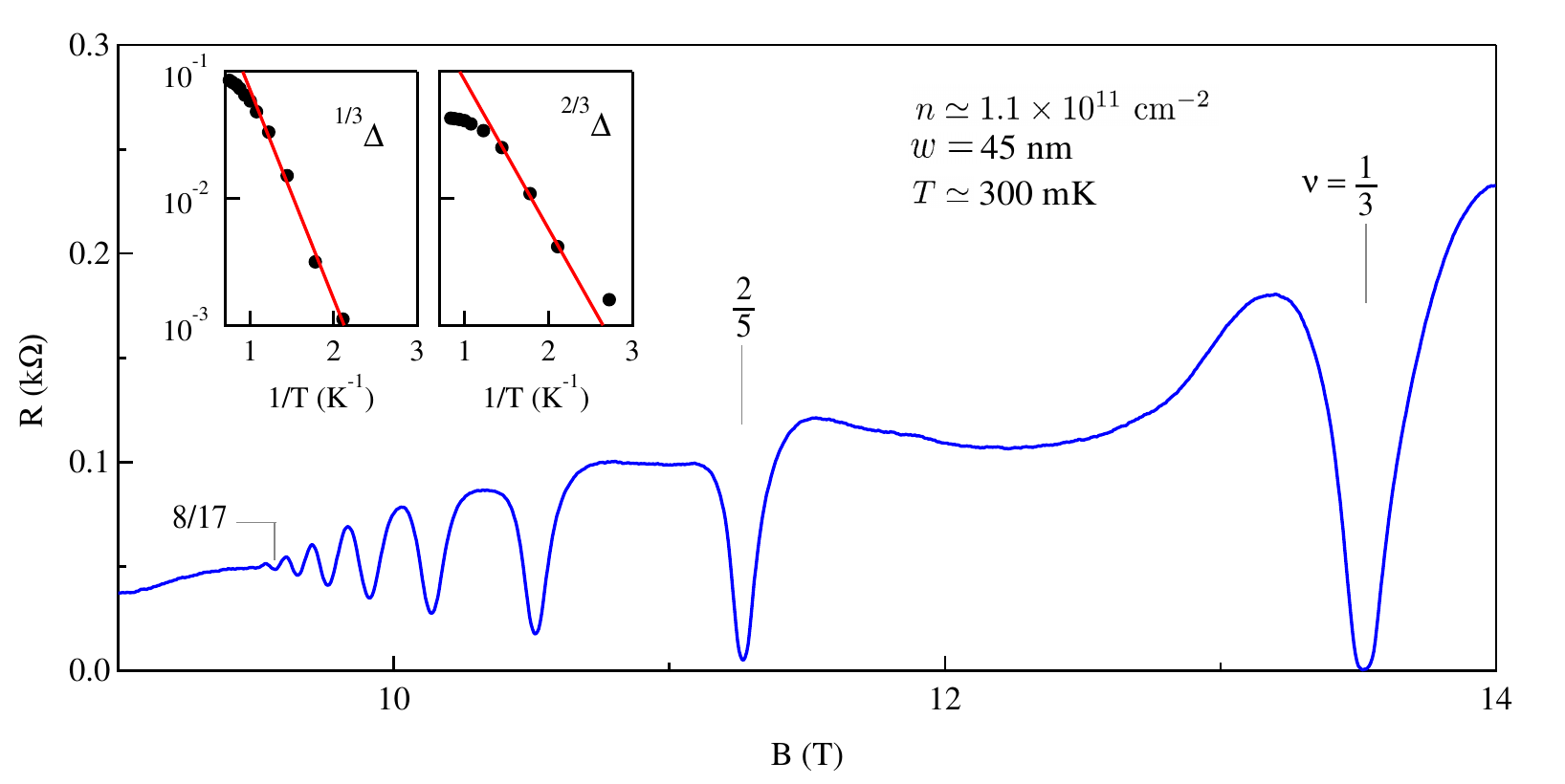, width=1\textwidth}
  \centering
  \caption{\label{transport} 
   $R_{xx}$ vs. \textit{B} for a GaAs 2DES with $w=45$ nm. The deduced $^{1/3}\Delta$ and $^{2/3}\Delta$ are ($7.6$ $\pm$ $0.4$) K and ($5.3\pm0.3$) K, respectively.
  }
  \label{fig:transport}
\end{figure*}

\begin{figure}[h!]
  \centering
    \psfig{file=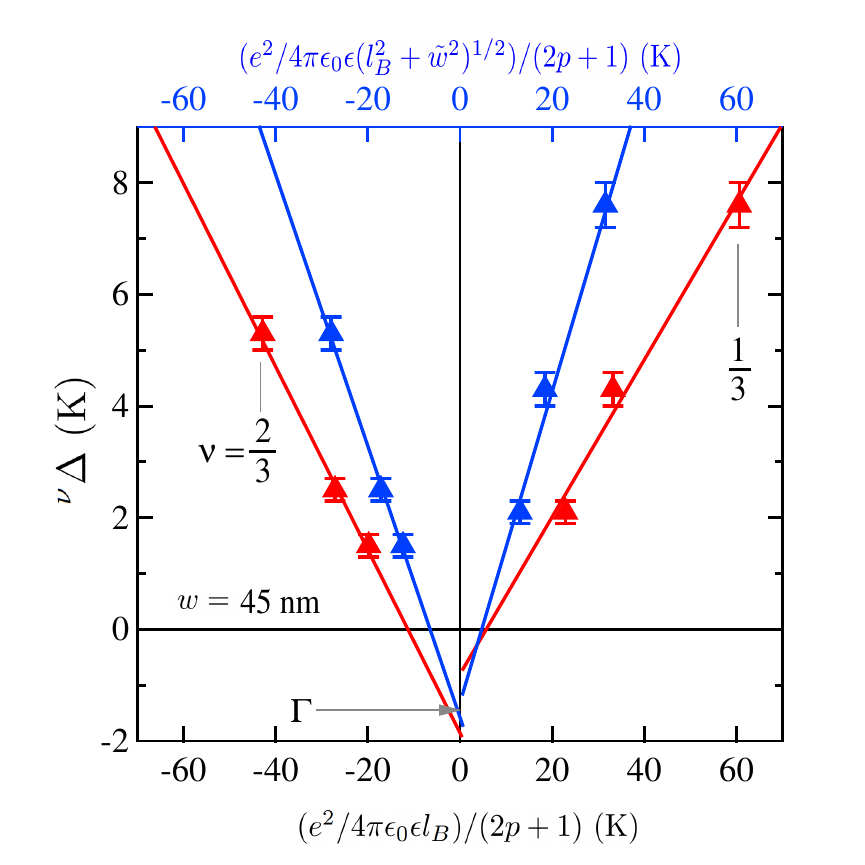, width=0.45\textwidth}
  \centering
  \caption{\label{transport} 
   Red symbols are $^{\nu}\Delta$ vs. $(e^{2}/4\pi\epsilon_{0}\epsilon l_{B})/(2p+1)$, for a 2DES with $w=45$ nm. The blue symbols are $^{\nu}\Delta$ vs. the Zhang-Das Sarma energy $(e^{2}/4\pi\epsilon_{0}\epsilon (l_{B}^{2}+\tilde w^{2})^{1/2})/(2p+1)$. The red and blue lines are linear fits to the data, and $\Gamma=1.5\pm0.5$ K.
  }
  \label{fig:transport}
\end{figure}

\clearpage

\begin{figure*}[h!]
  \centering
    \psfig{file=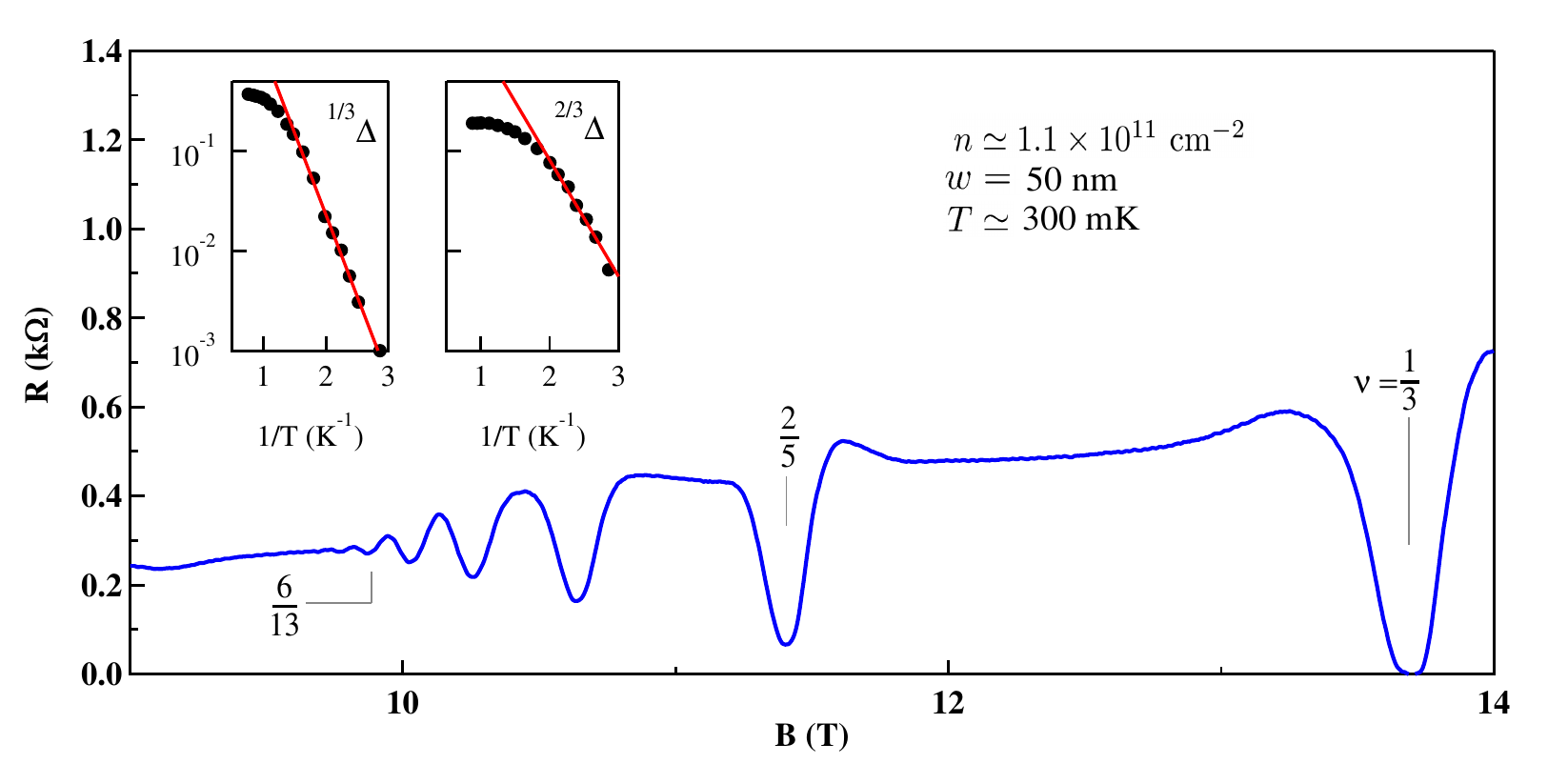, width=1\textwidth}
  \centering
  \caption{\label{transport} 
   $R_{xx}$ vs. \textit{B} for a GaAs 2DES with $w=50$ nm. The deduced $^{1/3}\Delta$ and $^{2/3}\Delta$ are ($7.5$ $\pm$ $0.2$) K and ($5.4\pm0.3$) K, respectively.
  }
  \label{fig:transport}
\end{figure*}

\begin{figure}[h!]
  \centering
    \psfig{file=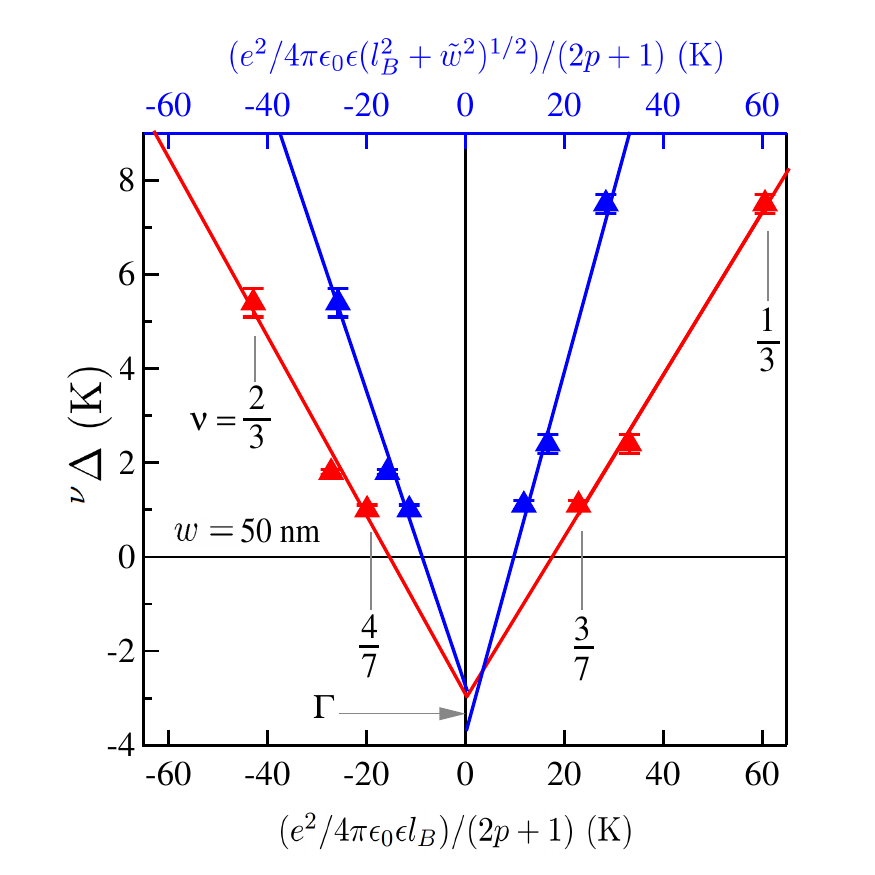, width=0.45\textwidth}
  \centering
  \caption{\label{transport} 
   Red symbols are $^{\nu}\Delta$ vs. $(e^{2}/4\pi\epsilon_{0}\epsilon l_{B})/(2p+1)$, for a 2DES with $w=50$ nm. The blue symbols are $^{\nu}\Delta$ vs. the Zhang-Das Sarma energy $(e^{2}/4\pi\epsilon_{0}\epsilon (l_{B}^{2}+\tilde w^{2})^{1/2})/(2p+1)$. The red and blue lines are linear fits to the data, and $\Gamma=3.3\pm0.6$ K.
  }
  \label{fig:transport}
\end{figure}

\clearpage

\begin{figure*}[h!]
  \centering
    \psfig{file=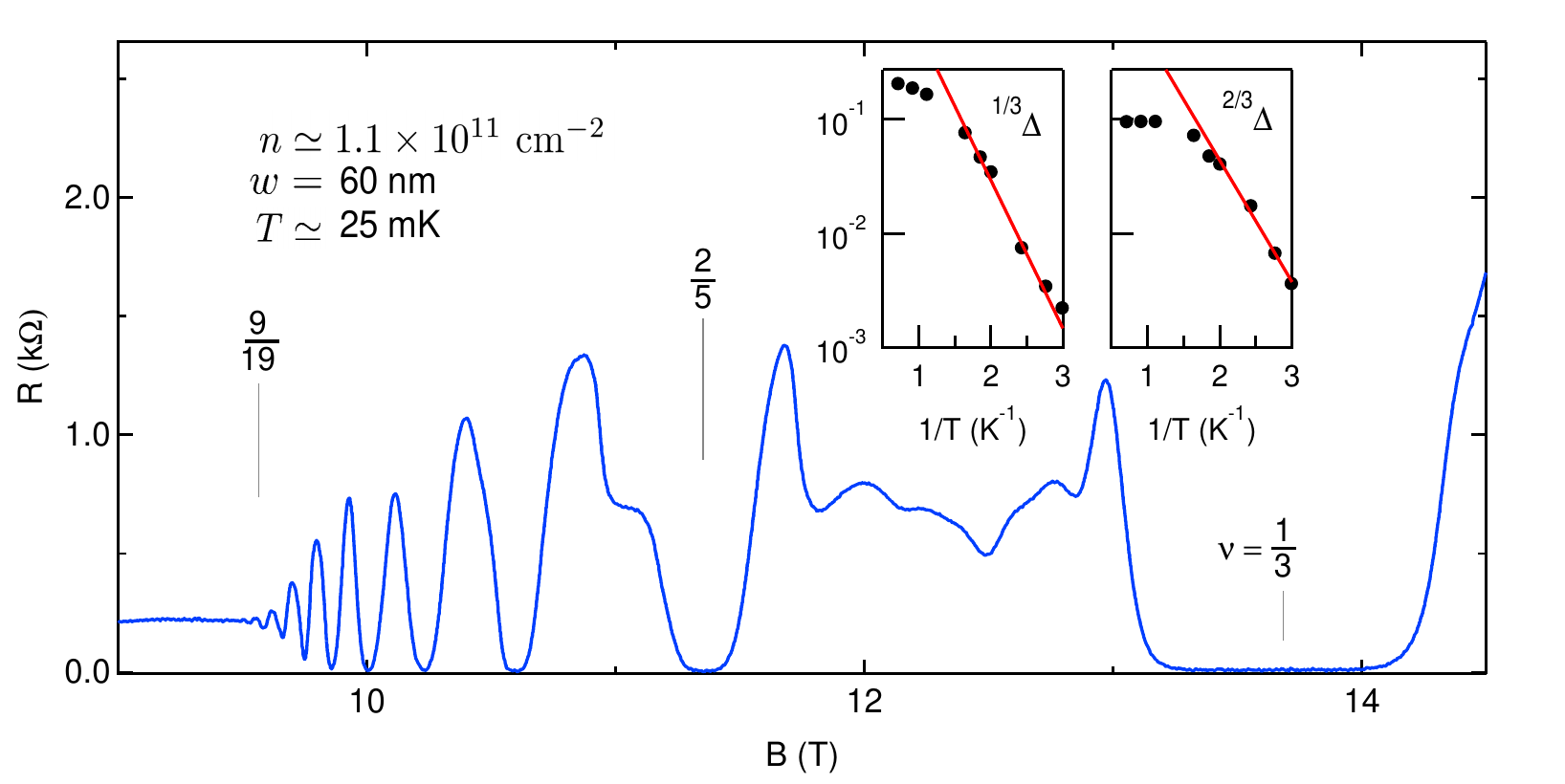, width=1\textwidth}
  \centering
  \caption{\label{transport} 
   $R_{xx}$ vs. \textit{B} for a GaAs 2DES with $w=60$ nm. The deduced $^{1/3}\Delta$ and $^{2/3}\Delta$ are ($6.0$ $\pm$ $0.5$) K and ($4.9\pm0.3$) K.
  }
  \label{fig:transport}
\end{figure*}

\begin{figure}[h!]
  \centering
    \psfig{file=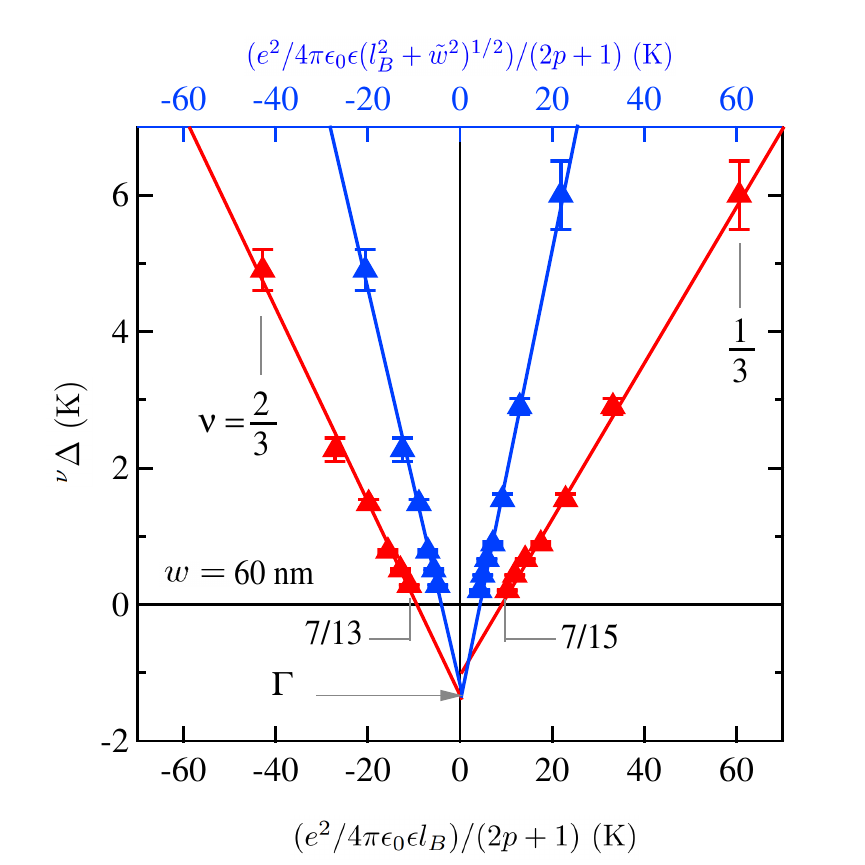, width=0.45\textwidth}
  \centering
  \caption{\label{transport} 
   Red symbols are $^{\nu}\Delta$ vs. $(e^{2}/4\pi\epsilon_{0}\epsilon l_{B})/(2p+1)$, for a 2DES with $w=60$ nm. The blue symbols are $^{\nu}\Delta$ vs. the Zhang-Das Sarma energy $(e^{2}/4\pi\epsilon_{0}\epsilon (l_{B}^{2}+\tilde w^{2})^{1/2})/(2p+1)$. The red and blue lines are linear fits to the data, and $\Gamma=1.3\pm0.1$ K.
  }
  \label{fig:transport}
\end{figure}

\clearpage

\begin{figure*}[h!]
  \centering
    \psfig{file=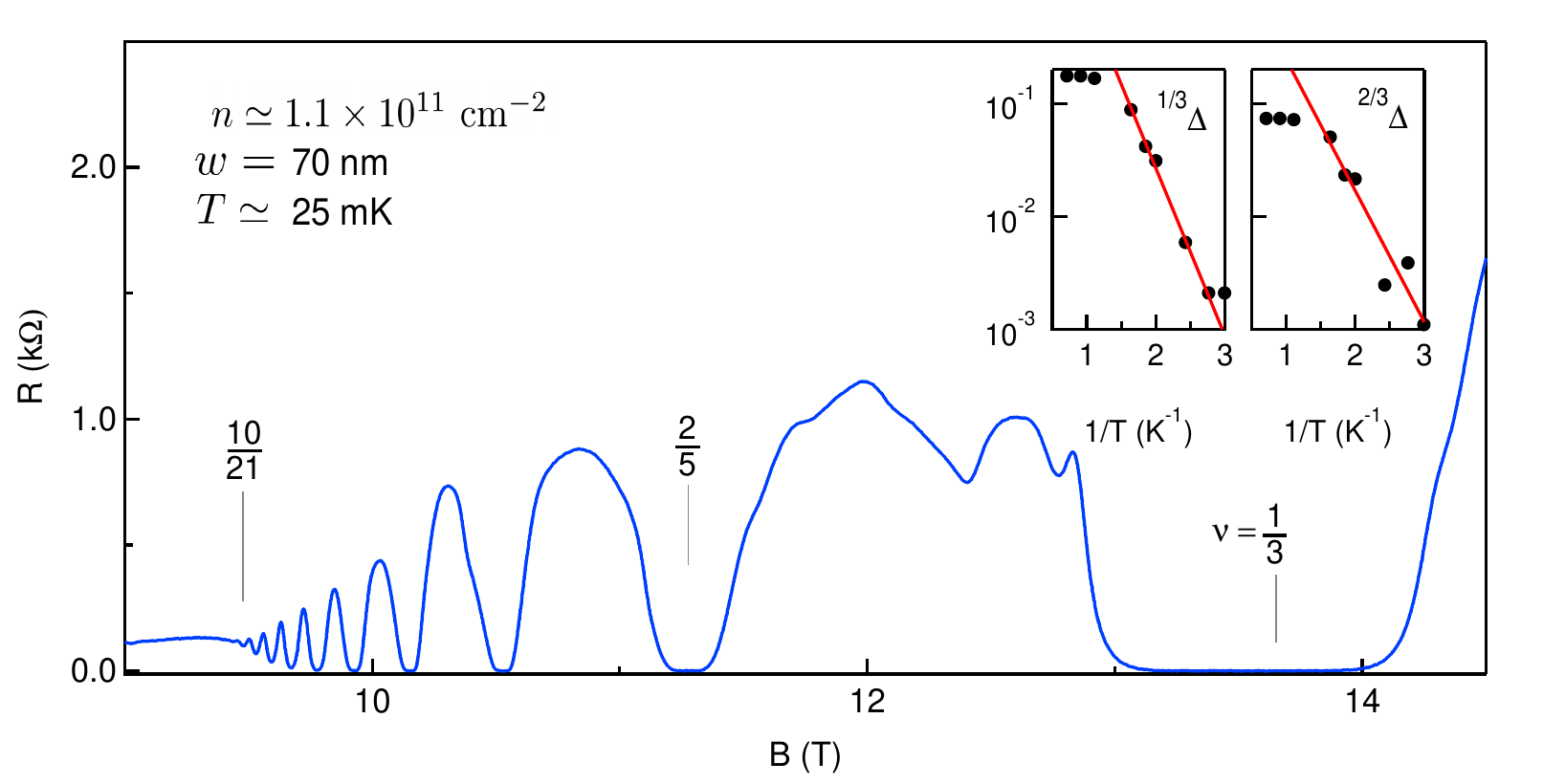, width=1\textwidth}
  \centering
  \caption{\label{transport} 
   $R_{xx}$ vs. \textit{B} for a GaAs 2DES with $w=70$ nm. The deduced $^{1/3}\Delta$ and $^{2/3}\Delta$ are ($6.8$ $\pm$ $0.4$) K and ($5.3\pm0.8$) K, respectively.
  }
  \label{fig:transport}
\end{figure*}

\begin{figure}[h!]
  \centering
    \psfig{file=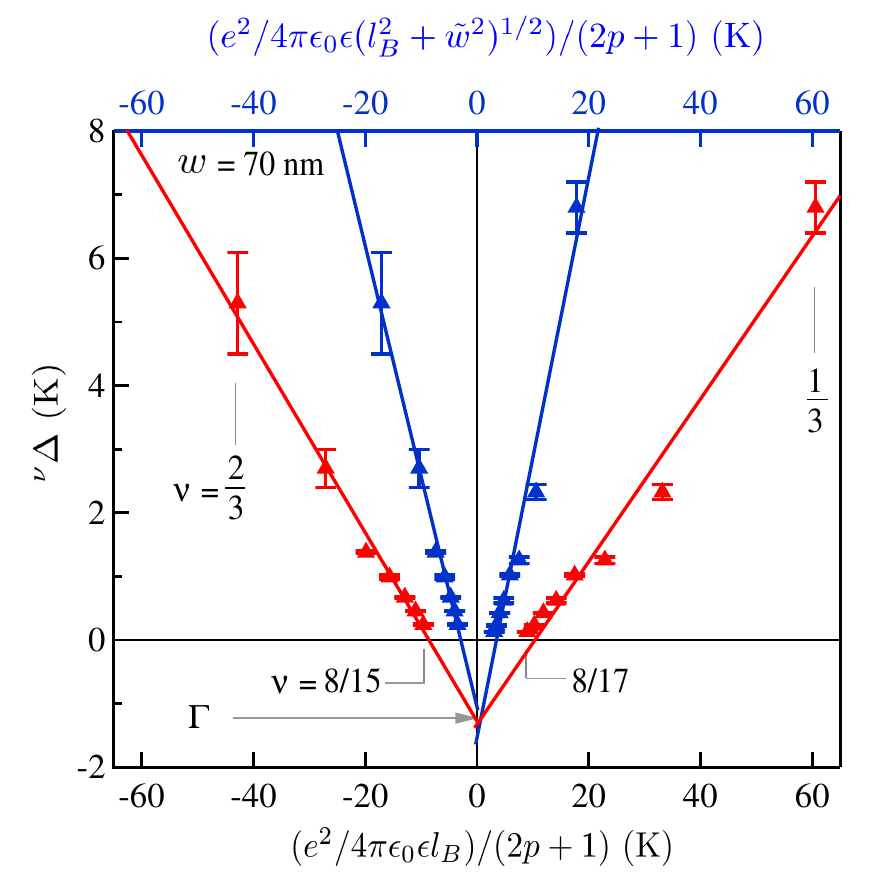, width=0.45\textwidth}
  \centering
  \caption{\label{transport} 
   Red symbols are $^{\nu}\Delta$ vs. $(e^{2}/4\pi\epsilon_{0}\epsilon l_{B})/(2p+1)$, for a 2DES with $w=70$ nm. The blue symbols are $^{\nu}\Delta$ vs. the Zhang-Das Sarma energy $(e^{2}/4\pi\epsilon_{0}\epsilon (l_{B}^{2}+\tilde w^{2})^{1/2})/(2p+1)$. The red and blue lines are linear fits to the data, and $\Gamma=1.3\pm0.2$ K.
  }
  \label{fig:transport}
\end{figure}

\clearpage

\section{Energy gaps at $\nu=1/3$, $2/3$, $2/5$, and $3/5$}

Figure S15(a) displays $^{\nu}\Delta$ (in Kelvin) vs. quantum well width $w$ for $\nu=1/3$ and $\nu=2/3$. Figure S15 (b) presents $^{\nu}\Delta$ (in units of Coulomb energy, $^{\nu}E_{C}$) vs. effective electron layer thickness $\tilde w$ (in units of the magnetic length, $l_{B}$) for $\nu=1/3$ and $2/3$. All $^{\nu}\Delta$ fall below $0.06E_{C}$, on the other hand, the theoretical value of $^{1/3}\Delta$ and $^{2/3}\Delta$ is $0.10E_{C}$ for an ideal 2DES, i.e., no Landau level mixing, zero electron layer thickness, and no disorder. Qualitatively, all energy gaps decrease as $\tilde w$ increases in accordance with the softening of the Coulomb interaction for 2DESs with large electron layer thicknesses. 

Figure S16 presents the energy gaps with the disorder parameter ($\Gamma$) added. Figure S16(a) shows $(^{2/3}\Delta+\Gamma)$ vs. $\tilde w$; the disorder parameter is displayed in Fig. S16(b). Note that, $\Gamma$ in units of $^{2/3}E_{C}$ can be as large as one third of the gaps reported in Fig. S16(a). As in the main text, the measured energy gaps approach the calculated values for large electron layer thicknesses.  To summarize our experimental data, we show $(^{\nu}\Delta+\Gamma)$ for $\nu=1/3$ and $2/3$ in Fig. S16(c). We find reasonable agreement with calculations for large electron layer thicknesses, but the experimental data points deviate from the calculated predictions for $\tilde w\leq 1$. 

Figure S17 presents energy gaps for $\nu=2/5$ and $3/5$.  The magnitude of the energy gap for $\nu=2/5$ and $3/5$, in an ideal 2DES, is $0.06E_{C}$, and all of our measured $^{\nu}\Delta$ fall bellow this value. In Fig. S18 (a) and (c), we show the energy gaps once disorder is taken into account experimentally for $\nu=2/5$ and $3/5$. For $\tilde w\geq 1$ the measured data points are in reasonable agreement with the available calculations (Ref. [25] in the main text). Three remarks are in order. First, $^{\nu}\Delta$ for $\nu=2/5$ and $3/5$ are smaller in magnitude compared to $^{1/3}\Delta$ and $^{2/3}\Delta$. Second, the $\Gamma$ values can be as large as $50\%$ of $^{3/5}\Delta$ (for $S_{50}$, $\Gamma$ has the same value as $^{\nu}\Delta$). Thus, $\Gamma$ can have a larger impact on the behavior of $^{\nu}\Delta$ for $\nu=2/5$ and $3/5$. Figures S18 (b) and (d) show $\Gamma$ (in units of $^{\nu}E_{C}$) for $\nu=2/5$ and $\nu=3/5$, respectively. Finally for completeness, we plot $^{\nu}\Delta+\Gamma$ vs. $\tilde w$ for $\nu=2/5$ and $3/5$ in Fig. S19. As in Fig. S16(c), there is an overall reasonable agreement between the experimental data and theory, but there are some noticeable discrepancies.

\clearpage

\begin{figure*}[h!]
  \centering
    \psfig{file=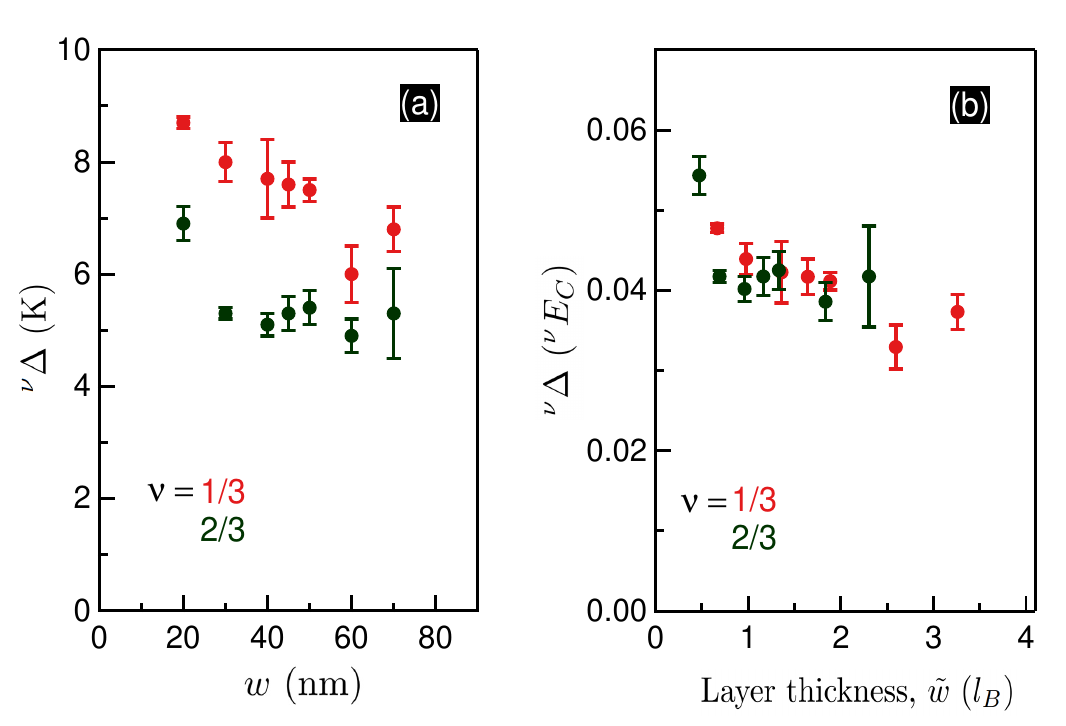, width=0.7\textwidth}
  \centering
  \caption{\label{transport} 
   (a) $^{\nu}\Delta$ (in Kelvin) vs. $w$.  (b) $^{\nu}\Delta$ (in units of $^{\nu}E_{C}$) vs. layer thickness, $\tilde w$ of $\nu=1/3$ and $2/3$. The values of $l_{B}$ are $7.1$ and $10.0$ nm for $\nu=1/3$ and $2/3$, respectively.
  }
  \label{fig:transport}
\end{figure*}

\begin{figure*}[h!]
  \centering
    \psfig{file=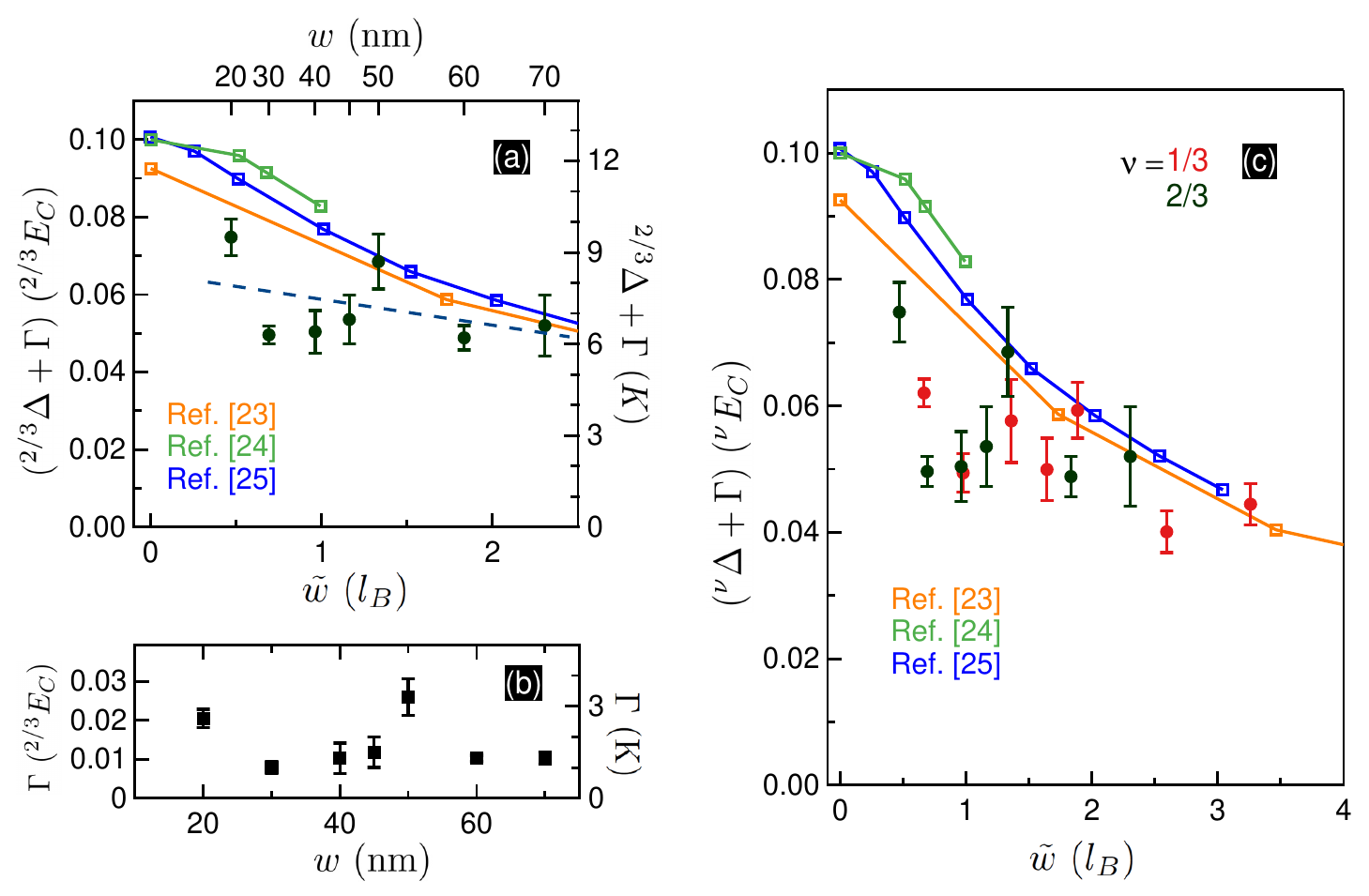, width=0.8\textwidth}
  \centering
  \caption{\label{transport} 
   (a) $(^{2/3}\Delta+\Gamma)$ vs. $\tilde w$. (b) $\Gamma$ vs. $\tilde w$. (c) $(^{\nu}\Delta+\Gamma)$ vs. $\tilde w$ for $\nu=1/3$ and $2/3$.
  }
  \label{fig:transport}
\end{figure*}

\clearpage

\begin{figure*}[h!]
  \centering
    \psfig{file=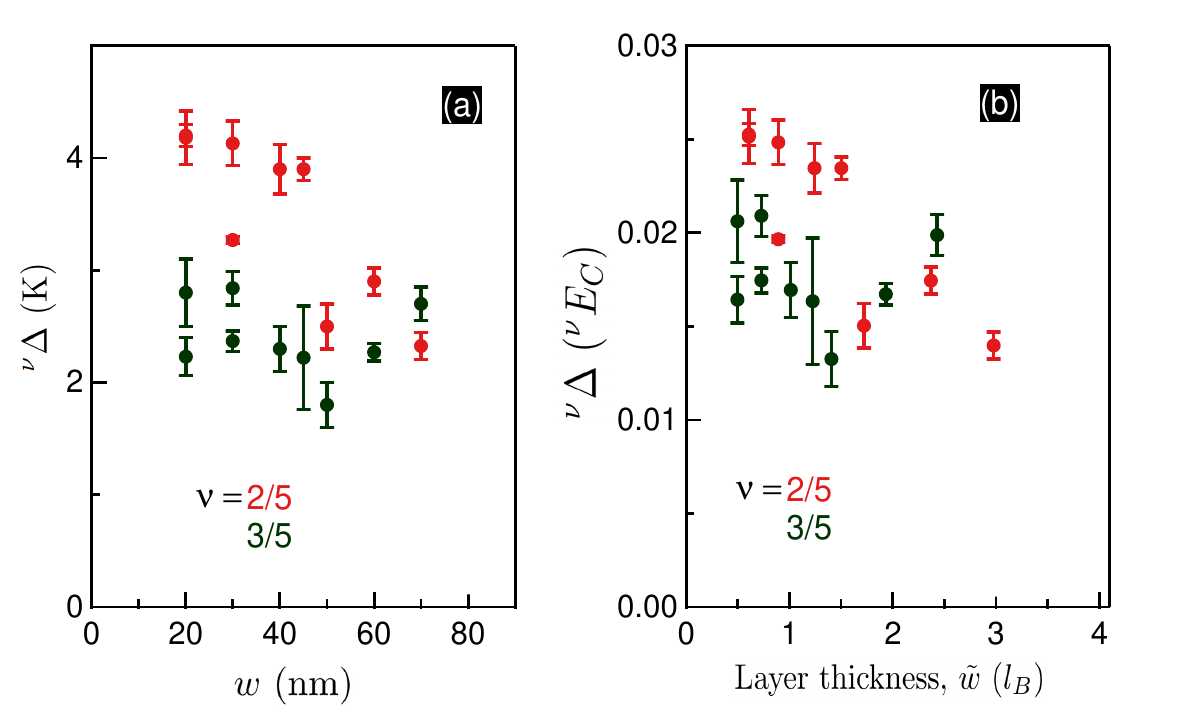, width=0.7\textwidth}
  \centering
  \caption{\label{transport} 
   (a) $^{\nu}\Delta$ (in Kelvin) vs. $w$. (b) $^{\nu}\Delta$ (in units of $^{\nu}E_{C}$) vs. layer thickness, $\tilde w$ of $\nu=2/5$ and $3/5$. The values of $l_{B}$ are $7.8$ and $9.5$ nm for $\nu=2/5$ and $3/5$, respectively. 
  }
  \label{fig:transport}
\end{figure*}

\begin{figure*}[h!]
  \centering
    \psfig{file=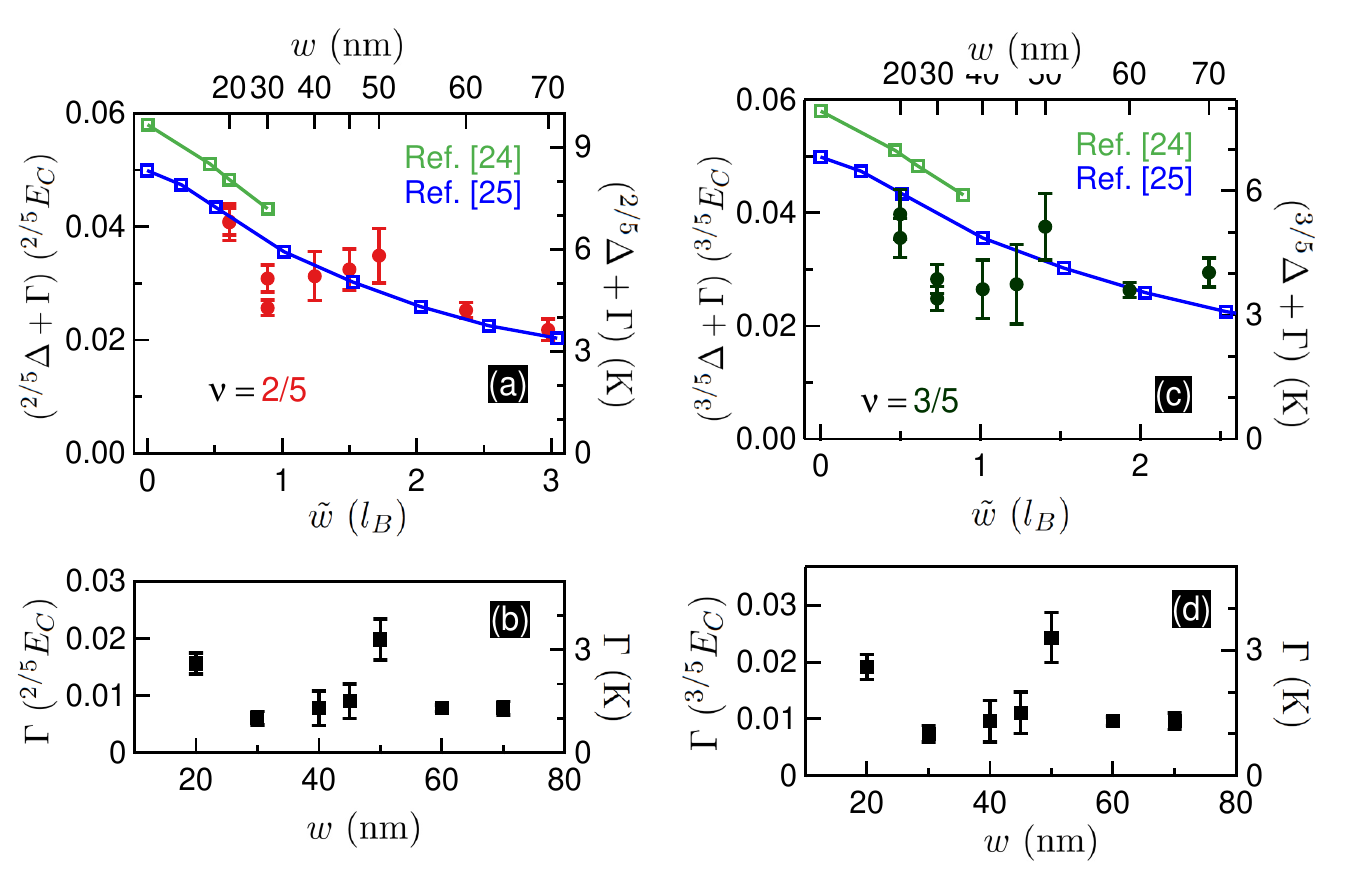, width=0.9\textwidth}
  \centering
  \caption{\label{transport} 
   (a) $(^{2/5}\Delta+\Gamma)$ vs. $\tilde w$ for $\nu=2/5$. (b) $\Gamma$ vs. $\tilde w$ for $\nu=2/5$. (c) $(^{3/5}\Delta+\Gamma)$ vs. $\tilde w$ for $\nu=3/5$ (d) $\Gamma$ vs. $\tilde w$ for $\nu=3/5$.
  }
  \label{fig:transport}
\end{figure*}

\clearpage

\begin{figure*}[h!]
  \centering
    \psfig{file=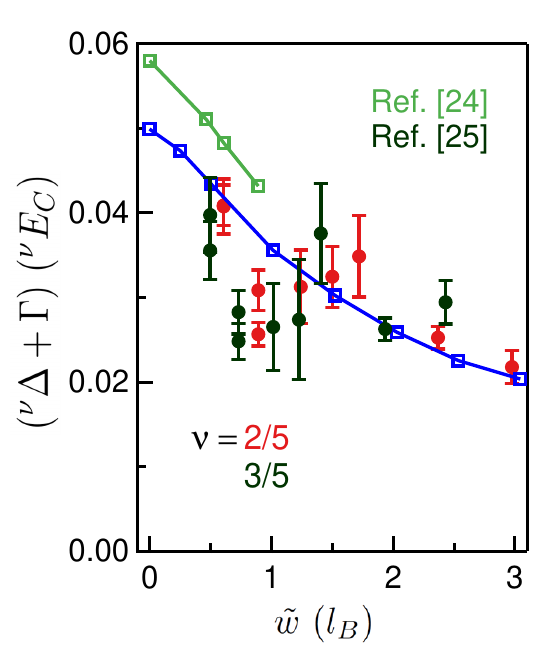, width=0.5\textwidth}
  \centering
  \caption{\label{transport} 
   $(^{\nu}\Delta+\Gamma)$ vs. $\tilde w$ for $\nu=2/5$ and $3/5$.
  }
  \label{fig:transport}
\end{figure*}